\documentclass[review]{elsarticle}
\usepackage[table]{xcolor}
\usepackage{comment}
\usepackage{xcolor}
\usepackage{epsfig}
\usepackage{subcaption}
\usepackage{calc}
\usepackage{url}
\usepackage{amssymb}
\usepackage{amstext}
\usepackage{amsmath}
\usepackage{amsthm}
\usepackage{array}
\usepackage{makecell}
\usepackage{graphicx}
\usepackage{wasysym}
\usepackage{amssymb}% http://ctan.org/pkg/amssymb
\usepackage{pifont}% http://ctan.org/pkg/pifont
\usepackage{xspace}
\usepackage{hhline}
\usepackage{multirow}
\usepackage{multicol}
\usepackage{booktabs}
\usepackage{lineno,hyperref}
\usepackage{adjustbox}
\modulolinenumbers[5]
\usepackage{stfloats}
\usepackage{rotating}
\usepackage{caption}
\usepackage{multirow,multicol, makecell, booktabs}
\theoremstyle{definition}
\usepackage{mdframed}
\usepackage{cleveref}
\usepackage[most]{tcolorbox}
\newtcbtheorem[]{Example}{\emph{Example .}}{
  enhanced,
  sharp corners,
  attach boxed title to top left={
    yshifttext=-1mm
  },
  colback=white,
  colframe=black!45!gray,
  fonttitle=\bfseries,
  boxed title style={
    sharp corners,
    size=small,
    colback=gray!45!black,
    colframe=black!45!black,
  } 
}{thm}

\journal{Journal of Computers and Security}

\usepackage{amsfonts}

\begin{document}

\begin{frontmatter}
% Elsevier \LaTeX\ template\tnoteref{mytitlenote}
\title{Identifying and Quantifying Trade-offs in Multi-Stakeholder Risk Evaluation with Applications to the Data Protection Impact Assessment of the GDPR}
% \tnotetext[mytitlenote]{Fully documented templates are available in the elsarticle package on \href{http://www.ctan.org/tex-archive/macros/latex/contrib/elsarticle}{CTAN}.}

\author[label1,label2]{Majid Mollaeefar}
\author[label1,label3]{Silvio Ranise}

\address[label1]{FBK-Center for Cybersecurity, Trento, Italy}
\address[label2]{DIBRIS, University of Genova, Italy}
\address[label3]{Department of Mathematics, University of Trento, Trento, Italy}

\address{\{mmollaeefar, ranise\}@fbk.eu}

\begin{abstract}
 Cybersecurity risk management consists of several steps including the selection of appropriate controls to minimize risks.  
This is a difficult task that requires to search through all possible subsets of a set of available controls and identify those that minimize the risks of all stakeholders.
Since stakeholders may have different perceptions of the risks (especially when considering the impact of threats), conflicting goals may arise that require to find the best possible trade-offs among the various needs.
In this work, we propose a quantitative and (semi-)automated approach to solve this problem based on the well-known notion of Pareto optimality.  For validation, we show how a prototype tool based on our approach can assist in the Data Protection Impact Assessment mandated by the General Data Protection Regulation on a simplified---but realistic---use case scenario.  We also evaluate the scalability of the approach by conducting an experimental evaluation with the prototype with encouraging results.

\end{abstract}

\begin{keyword}
%Cybersecurity, Decision Making, GDPR, 
Data Protection Impact Assessment, GDPR, Multi-Stakeholder Risk Assessment, Multi-Objective Optimization, Pareto-Optimality 
% \texttt{elsarticle.cls}\sep \LaTeX\sep Elsevier \sep template
% \MSC[2010] 00-01\sep  99-00
\end{keyword}

\end{frontmatter}

%\linenumbers
\section{Introduction}

%As technology develops rapidly, cyberspace shifts and digitalization increases, organizations may face growing risks of cybersecurity threats that might have a negative impact on their organization and their business intentions.
% Thus the management of these cybersecurity risks is crucial for organizations.
Cybersecurity risk management, i.e.\ the identification, evaluation, and prioritization of risks followed by the application of controls to minimize cyber risks,  is a vital aspect of the risk management process of any organization. 
%By performing a cybersecurity risk assessment, the organization should be able to
%\begin{itemize}[(i)]
%\item identify threats, i.e.\ those events that may cause “something to go wrong,” that are typically the outcome of malicious acts and might have undesirable business impacts and 
%\item determine the risk exposure levels, where a clear understanding of these risk levels would allow the organization to devote adequate action and resources to alleviate risks.
%\end{itemize}
% Thus the management of these cybersecurity threats is crucial for organizations. Cybersecurity risk assessment is a vital aspect of the risk management process of any organization. By performing a cybersecurity risk assessment, the organization would be capable to; (i) identify “what could go wrong” events (threat events) that are typically the outcome of malicious acts and that might have undesirable business impacts. (ii) determine the risk exposure levels, where a clear understanding of these risk levels would allow the organization to devote adequate action and resources to alleviate risks.
Several approaches are available to identify, evaluate, and prioritize cybersecurity threats such as the NIST Risk Management Framework\footnote{\url{https://csrc.nist.gov/projects/risk-management/}} that consists of several steps including the selection of controls necessary to protect the system and organization commensurate with risk.  This is a non-trivial task as it typically requires to (a) search through a large space of possible configurations for controls mitigating a set of threats according to (b) how the various stakeholders (e.g., the organization providing a service and the users using it) perceive risks. Different attitudes to risk by the various stakeholders may give rise to conflicting goals when considering additional constraints such as costs and skills required to deploy controls; for instance, customers of an online banking service may be interested to eliminate all threats for their financial transactions while the bank is willing to provide protection for the most common vulnerabilities while accepting the risk of more sophisticated attacks to maintain costs at an acceptable level.  In this paper, we consider the problem of providing automated assistance to the process of selecting the best possible configurations of controls to mitigate risks for all the stakeholders by making the following three contributions:
\begin{enumerate}[(C1)]
    \item we describe a methodology to semi-automatically assist stakeholders in the definition of their objectives that measure how much risks are reduced by adopting a certain configuration of the controls (this addresses point (b) above and is done by extracting crucial information already elicited during the application of the adopted approach to risk management);
    \item we define a decidable multi-objective optimization problem (based on the objectives previously identified)---called Multi-Stakeholder Risk Minimization Problem (MSRMP)---whose Pareto optimal solutions (see, e.g., \cite{marler2004survey}) are the subsets of the controls for which no stakeholder's risk can be further reduced without increasing the risk of at least one of the other stakeholders (this is a first step towards addressing point (a) above and is done by exploiting automated state-of-the-art tools for computing the set of solutions);
    \item designing and experimentally evaluating heuristics to visit the set of all possible configurations and guarantee the scalability of the proposed technique (this complements the previous contribution to address point (a) by identifying appropriate strategies to partition large search spaces to make the approach viable in practice).
\end{enumerate}
The ability to tackle this kind of problem is particularly relevant when considering privacy provisions deriving from national or international regulations.  For instance, the General Data Protection Regulation (GDPR)~\cite{regulation2016} requires to conduct a Data Protection Impact Assessment (DPIA) to guarantee the protection of personal data and preserve the rights and freedom of individuals.  This means that the organization offering a data processing activity should reduce the risk of the user to an acceptable level while controlling costs and other business goals.  In this context, being able to compute the subsets of controls that minimize the risks of both the organization of the system and  its users is a necessary pre-requisite to identify the most appropriate configuration of the controls that offer the best possible trade-off among the various objectives.    

\paragraph{Plan of the paper}  
In Section~\ref{sec:MSRToAP}, we introduce the Multi-Stakeholder Risk Minimization Problem (MSRMP) and its formalization as a multi-objective optimization problem (cf.\ contribution (C2) above) together with an approach to reduce the search space (cf.\ contribution (C3) above).  For concreteness, we propose a running example to illustrate the main ideas underlying the problem (Section~\ref{subsec:scenario}).  To find all Pareto optimal solutions and assist stakeholders to identify the risk management policies under which the risk exposure is minimized,  we propose an automated technique to solve MSRMP instances (Section~\ref{subsec:problem}). In Section~\ref{sec:new-methodology}, we discuss a methodology to assist stakeholders in the definition of instances of the MSRMP (cf.\ contribution (C1) above).  In Section~\ref{sec:impl-exp}, we describe a tool supporting the definition of MSRMP instances and the computation of their solutions together with a set of experiments aiming to understand the effectiveness of the strategies to reduce the search space and thus improving the scalability of the proposed approach (cf.\ contribution (C3) above). We discuss related works (Section~\ref{sec:relatedwork}) and conclude the paper with a summary of the main contributions and some hints for future work (Section~\ref{sec:conclusion}).

\section{Multi-Stakeholder Risk Minimization %Trade-off Analysis 
Problem (MSRMP)}\label{sec:MSRToAP}
%In a multi-stakeholder risk assessment, different stakeholders have different preferences to assess the potential impact of threats. We define a risk management policy (RMP) as a set of technical and organizational measures that must put in place to deal with risk. 
%The purpose of a RMP is to minimize the risks for all the considered stakeholders while considering all constraints, including legal prescriptions and business requirements.

Cyber-risk is a measure of the likelihood and the impact of threats, i.e.\ circumstances or events with the potential to harm a cyber-system such as  the unauthorized disclosure, destruction, modification, or interruption of system assets.  Cyber-risk management is the \emph{identification} and \emph{assessment} of risks followed by the \emph{definition} and \emph{enforcement} of appropriate \emph{mitigation measures} for risk minimization. 
% We briefly discuss each step.
The identification of risks depends on the assets of the system to be protected and requires to perform threat modeling, i.e.\ to understand and describe how an adversary might compromise a system.  The assessment of risks amounts to evaluating the impact and the likelihood of the various threats.  
For instance, a backdoor in a certain version of an operating system may have a dramatic impact.  The risk may be severe if patches are applied late as the likelihood that an adversary exploits the vulnerability is high whereas the risk becomes small when patches are quickly applied as the time-window during which an attacker can exploit the vulnerability is substantially reduced.  The balance between impact and likelihood is key to risk assessment.  Once risks have been identified and assessed, suitable Risk  Management Policies (RMPs) should be defined and enforced.  RMPs comprise both technical (e.g., deploy the latest version of the Transport Layer Security protocol) and organizational (e.g., a cyber security awareness and training program for employees) measures to minimize risks.  Indeed, the ultimate goal of risk management is to minimize risks while maximizing the chances to reach business objectives and complying with legal provisions, such as the GDPR.  Indeed, failing to do this may bring in additional risks and costs due to an unsatisfactory return on investment or fines for lack of compliance.  

Given the increasing complexity of cyber-systems, it is routine that several stakeholders cooperate in their design, development, and deployment.  This further complicates risk management.  For instance, according to the GDPR, in case a system processes personal data, its data controller 
% (i.e.\ the natural or legal person, public authority, agency or other body which, alone or jointly with others, determines the purposes and means of the processing of personal data)
shall guarantee that the risk of violating the rights and freedom of the data subjects 
% (i.e.\ the users from whom or about whom a data controller collects personal information)
is low. The data controller must do this by considering state-of-the-art RMPs and budget constraints.  When the data controller involves a data processor,  
% (i.e.\ an entity processing personal data on behalf of the data controller)
the latter may have strict computational constraints for scalability and efficiency that, in turn, guarantee economy of scale.  While the various stakeholders may agree on a common set of threats for a given system together with their likelihood, they will have diverging criteria to evaluate the potential impact of the identified threats.
For instance, data subjects will favor comprehensive RMPs to reduce the risk of data breaches. In contrast, a data controller or a data processor may be more interested in cheap and easy to enforce RMPs that cover most threats while neglecting those less likely to occur.
% For instance, data subjects will favour comprehensive RMPs to reduce the risk of data breaches while a data controller or a data processor may be more interested in cheap and easy to enforce RMPs that cover most threats while neglecting those that are less likely to occur.  
Besides making the definition of the impact of threats dependent on each stakeholder, this greatly complicates the search for RMPs that minimize risks.  Indeed, the search for RMPs that simultaneously minimize the risk level for each stakeholder becomes a non-trivial task in the presence of conflicting objectives and requires the adoption of the notion of Pareto optimality. 
To understand the problem, consider the situation in which we have two RMPs  $\mathit{rpm1}$ and $\mathit{rpm2}$ with risk vectors $\langle 1,2,1 \rangle$ and $\langle 1,1,2 \rangle$, respectively, where the first component is the risk of the data subject, the second is that of the data controller, and the third is that of the data processor.  The data subject has no preference between the two RMPs, the data controller prefers $\mathit{rpm1}$ over $\mathit{rpm2}$, and the data processor $\mathit{rpm2}$ over $\mathit{rpm1}$.  In other words, no RMP minimizes the risk for all the stakeholders; so, which one between $\mathit{rpm1}$ over $\mathit{rpm2}$ should be preferred? 
According to the notion of Pareto optimality (see, e.g.,~\cite{marler2004survey}),
% The answer can be given by using the notion of Pareto optimality as it has been introduced in Section~\ref{subsec:moop};
% (see,  e.g.,~\cite{marler2004survey}); 
% in our context, this means that an RMP is Pareto optimal if there does not exist another RMP that improves the risk exposure for one stakeholder without detriment to that of another one.  In the example above, 
both $\mathit{rpm1}$ and $\mathit{rpm2}$ are to be considered optimal and further aspects need to be considered to select one of the two such as the fact that one of the two promises to provide a higher return on investment or that it is easier to show its compliance with the GDPR or other legal provisions.  Because vectors cannot be ordered completely, all the Pareto optimal solutions can be regarded as equally desirable in the mathematical sense and
we need a decision maker to select the preferred one among
them. To enable the decision maker to do this, we need to be able to compute the set of Pareto optimal solutions. Below (Section~\ref{subsec:problem}), we formalize the problem of finding Pareto optimal configurations of RMPs, i.e., configurations minimizing the risk of stakeholders, 
in the framework of multi objective optimization and show how it can be solved by using general purpose algorithms under reasonable assumptions. 
Preliminary, we introduce a simplified but realistic running example to better grasp the problem.
% We also use it to illustrate the concepts introduced in the paper.
%Risk  Management  Policies (RMPs)---intended as the set of technical and organizational measures put in place to monitor, control, and minimize risks---have different effects on the risk exposure of each stakeholder. A RMP should therefore be selected with the goal of minimizing the risks for all the considered stakeholders while considering all possible constraints, such as legal prescriptions and business requirements. Indeed, failing to comply with applicable laws brings in additional risks and costs due to the expected fines and even if the law is not formally violated, unbalanced risk levels among stakeholders may point out potential conflicts of interests, which can result in generating additional risks. . 

\subsection{\textbf{Running Example: An Application of the GDPR's DPIA}}
\label{subsec:scenario}
% \begin{figure}[t]
% 	   \centering
% 		\includegraphics[width=\textwidth]{Figures/ACME-Scenario.pdf}
% 		\caption{Overview on the main stakeholders in the scenario and their interaction with the system's components.} \label{fig:scenario}
% \end{figure}
We consider the situation in which an Italian company, called ACME below for the sake of anonymity, that must perform a Data Protection Impact Assessment (DPIA) for one of its software applications, as required by Article 35 of the General Data Protection Regulation~\footnote{https://gdpr-info.eu/art-35-gdpr/} (GDPR).   
The goal of a DPIA is to protect the rights and freedom of EU citizens with particular relevance to those related to their privacy. For this, it is crucial to perform an appropriate privacy risk assessments.   There are three main stakeholders involved in the process, namely (i) the \emph{Data Subject}, the patient whose data are being collected, stored and processed by the application, (ii) the \emph{Data controller}, ACME which is responsible for offering the data processing activities implemented by the software application, and (iii) the \emph{Data processor}, a company mandated by the Data Controller to design and implement the application deploying the various data processing activities. The data processor is a third party organization, possibly external to the data controller.  In the rest of this section, we focus on the problem of identifying appropriate security controls among a set of available ones that minimize the risks of all three stakeholders.  A peculiarity of this risk assessment is that the data controller must perform it to make the risk of the data subjects acceptable.  Indeed, this may give rise to conflicts with the data controller's and data processor's requirements on budgets and skill's shortage.  

\begin{figure}[t]
	   \centering
		\includegraphics[width=\textwidth]{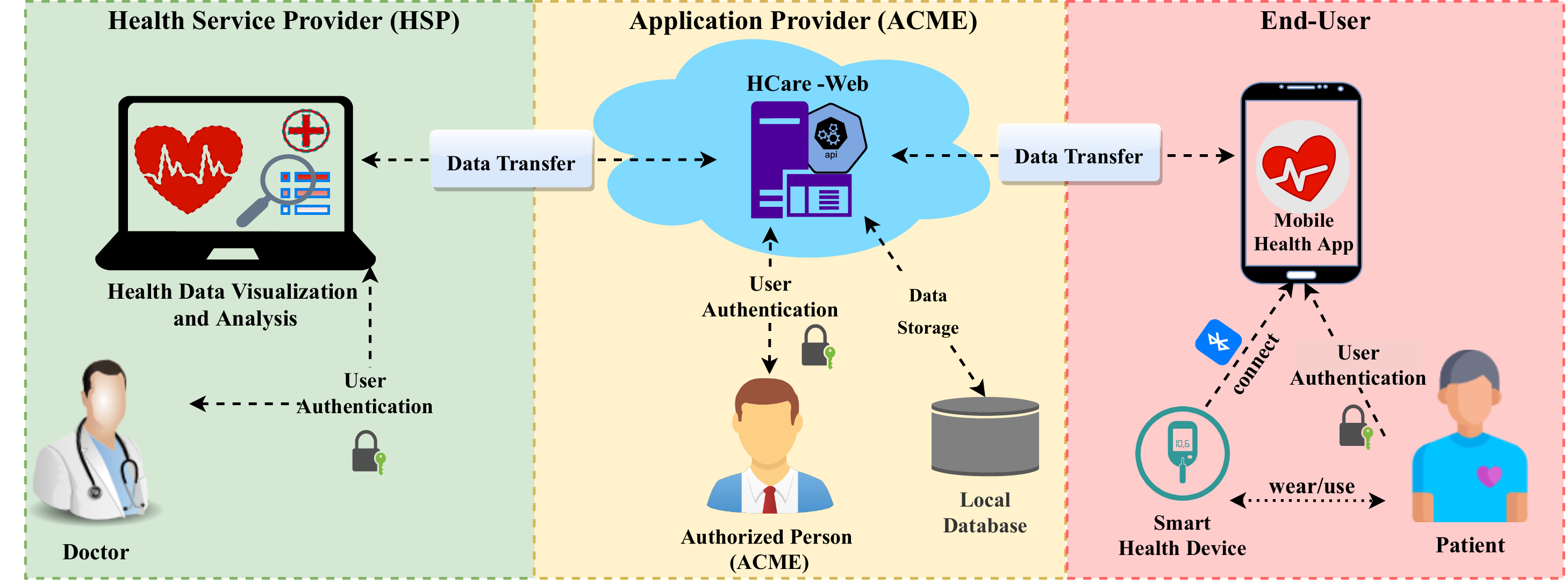}
		\caption{Overview on the main stakeholders in the scenario and their interaction with the system's components.} \label{fig:scenario}
\end{figure}

ACME % is the (anonymized) name of a real Italian startup operating in the healthcare domain. It 
develops a software application, called HCare, exposing an API service to allow its clients to work together, as illustrated in Figure \ref{fig:scenario}. 
Through the API, HCare connects three main stakeholders: the Health Service Provider (HSP), the API provider (ACME), and the patients which are the data controller, the data processor, and the data subjects in the context of the GDPR, respectively. 
Notice that an HSP in our case can also be an independent developer who provides IT-only services without offering actual health care support; for example, providing data visualization tools. Finally, the end-user is typically the patient using the app to send biometric data or user-initiated requests and receive responses from the HSP, e.g., prescriptions from a doctor, medical alerts, etc. 
HSPs use the APIs to perform some operations such as create, read, update, and delete (CRUD operations) in a compliant way – i.e., by considering proper roles and permissions and storing and accessing the data accordingly. 
The health data is stored in a cloud environment, controlled, and monitored by ACME.
Consequently, from a legal perspective, ACME acts as the data processor. However, due to the nature of its offered services, ACME has also to support data controllers to comply suitably.
Therefore, it looks at the issue of GDPR compliance from both perspectives, of the data processor and data controllers. This is handled by a service level agreement between ACME and the HSP.

ACME, as data controller, must be aware of how to properly process the patients' data because there could be a variety of harmful or threat events that could put even the patients' life at risk.
For instance, data (such as the patient's medical history) could be lost or corrupted due to a hardware failure. 
Patients may suffer severe consequences as a result of this situation because the healthcare data in question is used to offer healthcare services such as medical prescriptions, and missing or damaged data may result in incorrect diagnoses or the inability to provide the service.
For this reason, data storage must be trustworthy, which can be achieved by implementing appropriate data protection controls. 
For instance, more frequent backups or data replication are potential controls to mitigate possible risks in the case of a hardware failure.
However, these solutions change the risk exposure of ACME.
Data replication, particularly, introduces the needs and all the associated risks of a sophisticated network architecture. 
For example, business risks due to the rising costs, but also process risks due to the difficulty of network configuration.
This example demonstrates the consequences of the law: given that the data subject has certain fundamental rights, it is the data controller's responsibility to put in place the appropriate technical and organizational means to ensure that the rights of the data subject are respected.
The endeavor to reduce the risks for the data subject, on the other hand, may result in an increase in the risk exposure for ACME, which may include risks other than those related to personal data.
From these considerations we can see that it is likely that each stakeholder has different preferences for the various RMPs yielding different threat impact levels for each threat.
% We will determine these preferences for each stakeholder in Section~\ref{sec:impact-assessment}.
% The different threat impact levels cause the different risk exposure levels for each stakeholder. 
Therefore, we must solve the problem of selecting the optimal risk management policy, which we formalize in the framework of multi-objective optimization in the next section.

% PLEASE MAJID HAVE A LOOK AT THE FOLLOWING PARAGRAPH AND CHECK IF IT IS COHERENT WITH THE REST OF THE MANUSCRIPT... 

To summarize, for the running example, we consider a set $\mathcal{S}$ containing two stakeholders, namely the Data Controller and the Data Subject, a list of $5$ threats $T_1, ..., T_5$ shown in Table~\ref{tab:threat-table} (at page~\pageref{tab:threat-table}) and a list of associated security controls $c_1, ..., c_{25}$  shown in  (the first two columns of) Table~\ref{tab:threat-control} (at page~\pageref{tab:threat-control}).  Thus, we have $5$ threats and $25$ controls; the latter are associated to each threat as follows: $c_1, ..., c_5$ to $T_1$, $c_6, ..., c_{15}$ to $T_2$, $c_{16}, ..., c_{19}$ to $T_3$, $c_{20}, ..., c_{22}$ to $T_4$, and $c_{23}, ..., c_{25}$ to $T_5$.  In the next section, we use the running example to illustrate the formal notions we introduce albeit in a simplified form for the sake of simplicity and space.  So, for instance, we will consider only $3$ threats instead of $5$ and only $5$ security controls instead of $25$.  We observe that we use $c_1, ..., c_5$ as identifiers of the security controls in the following section for the sake of simplicity but they have been renamed in Table~\ref{tab:threat-control} where the whole set of controls is listed.  The solution of the multi-objective optimization problem in its full generality is discussed later in Section~\ref{subsec:appl-run-ex}.

% Figure~\ref{fig:scenario} depicts the main actors and their interactions with the system's components in our scenario. At the client-side (End-user), the Mobile Health App dynamically receives the biometric data from the smart health device, which is worn/used by the patient. Then the data is transferred to the HCare-web server (Application provider) for further processing. After receiving any updates (such as biometric data or user requests) from the patients, the doctors monitor their health condition at the health service provider side. Through the HCare-web, doctors can write a prescription, respond to requests, initiate a medical alert, etc. 

\subsection{\textbf{Problem Formalization}}
\label{subsec:problem}

Let $\mathcal{S}$ be a finite set of stakeholders and $\mathcal{T}$ a finite set of threats. For each stakeholder $s$ in $\mathcal{S}$, we assume a mapping $i_s:\mathcal{T}\to \mathcal{I}$ that computes the impact level of the harmful events generated by a threat $T$ when it occurs, where $\mathcal{I}$ is a sub-set of the reals denoting impact levels, intuitively $\mathit{il}_1<\mathit{il}_2$ implies that the impact level $\mathit{il}_1$ is less severe than the impact level $\mathit{il}_2$. 

 \begin{Example}{}{}
 \label{ex:start}
    \small {Referring to the example in Section~\ref{subsec:scenario}, the set $\mathcal{S}$ of stakeholders contains $s_1=$ \emph {Data controller (ACME)} and 
    $_2=$ \emph{Data subject (the patient)}.
    Consider the set $\mathcal{T}$ of threats to contain $T_1=$ \emph{Unlimited data storage}, $T_2=$ \emph{Unauthorized access}, and $T_3=$ \emph{Linkage attack}, as three potential threats. 
    We may define the mappings $i_{s_1}$ and $i_{s_2}$~$:\mathcal{T}\to \mathcal{I}$ by means of a table as follows:
    % Thus, we assume the mapping $i_s:\mathcal{T}\to \mathcal{I}$ is as bellow:
    % \\
    
    \begin{center}
    \begin{tabular}{c|c|c|c}{\centering}
   
    % $\mathcal{T}$ &$i_{s_1}$&$i_{s_2}$ \\ \hline\hline
    % $T_1$ & 0.6       & 0.2          \\ 
    % $T_2$ & 0.3       & 0.5         \\
    $~$          &$T_1$      &$T_2$ &$T_3$ \\ \hline\hline
    $i_{s_1}$ & 0.6       & 0.2  & 0.3         \\\hline 
    $i_{s_2}$ & 0.3       & 0.5  & 0.6         \\
    
    \end{tabular}
    \end{center}
    The values in the first and second rows of the table denote the impact levels for each threat from the point of view of the data controller ($s_1$) and data subject ($s_2$), respectively.  
    For instance, the  impact level associated to threat $T_1$ from the data controller point of view is $0.6$ whereas from the data subject point of view is $0.3$.}
    
 \end{Example}
 
As shown in the example above, $i_s$ is typically specified by using a tabular format. This is also the case for other mappings that we consider below.

% [[[I Added the following text in order to consider Preferences in the problem formalization ]]]

% Indeed, we associate to each stakeholder $s$ in $\mathcal{S}$, a finite set $\mathcal{P}_s$ of preferences that $s$ uses to evaluate the impact of the threats; for instance, the \emph{health condition} can be considered a data subject's preference where a hardware failure threat may remarkably impact it. 
% On the other hand, rising costs as a consequence of the threat may impact the \emph{business situation} (as a preference) of the data controller. 
% Formally, a preference in $\mathcal{P}_s$ is a mapping from $\mathcal{T}$ to a finite set values $\{0, 1,2,3, 4\}$ where $1$ means that the impact value is negligible whereas $4$ it is catastrophic. In some cases, a threat may have no impact on a preference which considers as $0$. 

Let $\mathcal{C}$ be a finite set of controls and $\{\mathcal{C}_T\}_{T\in \mathcal{T}}$ a family of finite set of controls; intuitively, $\mathcal{C}_T$ is the set of controls that, alone or in combination, may mitigate a threat $T$.  
\begin{Example}{}{}
   \small{ 
    
%   [[[THIS EXAMPLE MUST BE CHANGED ACCORDINGLY]]]
   To mitigate the risk of threats in Example 1, we identify a family of set of controls $\{\mathcal{C}_{T_1},\mathcal{C}_{T_2},\mathcal{C}_{T_3} \}$ where $\mathcal{C}_{T_1}= \{c_1,c_2\}$, $\mathcal{C}_{T_2}= \{c_3,c_4\}$, and $\mathcal{C}_{T_3}= \{c_5\}$. For instance, $c_1$ can be \emph{(Ensuring data minimization), $c_2$ (Enabling data deletion), $c_3$ (Ensuring secure storage), $c_4$ (Logging access to personal data), and $c_5$ (Ensuring data anonymization)}.
%   $\mathcal{C} =$ \{\emph{$c_1$ (Ensuring data minimization), $c_2$ (Enabling data deletion), $c_3$ (Ensuring secure storage), $c_4$ (Logging access to personal data), $c_5$ (Ensuring data anonymization)}\}\ 
%   where $c_1$ and $c_2$ are devoted to mitigate $T_1$, $c_3$ and $c_4$ to $T_2$, and $c_5$ to $T_3$.
   }
\end{Example}
% [[[PLEASE MAJID ADD AN EXAMPLE HERE]]]
% , e.g., for the hardware failure threat, implementing data replication and performing frequent backups are some possible controls. 
For each threat $T$ in $\mathcal{T}$, we assume a mapping $\mu_{T}:\mathcal{C}_T\to [0..1)$ that quantifies the mitigation by a control in $\mathcal{C}_T$ on the impact of a threat $T$.  Intuitively, $\mu_T(c)$ can have three possible statuses: (i) $\mu_T(c)=0$ clarifies that the control $c$ is not adopted and thus can not contribute in mitigating threat $T$, (ii) $0<\mu_T(c)<1$ means that the control $c$ is adopted and partially mitigates the threat $T$, and (iii) $\mu_T(c)=1$ represents that the control is adopted and fully mitigates $T$.
% Intuitively, $\mu_T(c)$ is the amplitude (it may be useful to think of it as a percentage) by which the impact of the threat $T$ is reduced with the adoption of the control $c\in \mathcal{C}_T$.  
% In many cases, the co-domain of $\mu_T$ is a finite subset of values in [0...1).

We are now in the position to define the impact residue of the threat $T$ under a given mitigation mapping $\mu_T$ as:
\begin{equation}
\label{eq:impact}
   \mathit{ir}_s(T)= i_s(T)\cdot (1-\frac{\Sigma_{c\in \mathcal{C}_T} \mu_T(c)}{|\mathcal{C}_T|})~ .
\end{equation} 
We observe that the expression between parentheses is the mitigation obtained by adopting some of the controls in $\mathcal{C}_T$ associated to $T$ and that the degree of effectiveness of a control $c$ in mitigating a threat $T$ is given by $\mu_T(c)$.  Because of its importance, we introduce the following abbreviation:  
\begin{equation}
\label{eq:mitigation-abbreviation}
   m(T)= \frac{\Sigma_{c\in \mathcal{C}_T} \mu_T(c)}{|\mathcal{C}_T|}~  
\end{equation} 
that depends on the mitigation mapping $\mu_T$ (and since $\mathit{ir}_s(T)=i_s(T)\cdot (1-m(T))$ also $\mathit{ir}_s(T)$ depends on $\mu_T$) but we avoid to make such a dependence explicit to simplify notation. Given a family $\{\mu_T\}_{T\in \mathcal{T}}$ of mitigation mappings, the overall impact residue for a given stakeholder $s\in \mathcal{S}$ is defined as $\mathit{oir}(s)= \Sigma_{T\in \mathcal{T}} \mathit{ir}_s(T)$, where $\mathit{ir}_s(T)$ is evaluated under the mitigation mapping $\mu_T$.  In other words, $\mathit{oir}(s)$ is the sum, over the set $\mathcal{T}$ of threats, of all impact residues, each one evaluated under the associated mitigation mapping in $\{\mu_T\}_{T\in \mathcal{T}}$.
%By introducing a variable $m(T)$ to replace the expression $\frac{\Sigma_{c\in \mathcal{C}_T} \mu_T(c)}{|\mathcal{C}_T|}$, we are able to rewrite (\ref{eq:impact}) as follows:
%\begin{equation*}
% \label{eq:impact}
%   $\mathit{ir}_s(T)= i_s(T)\cdot (1- m(T)).$
%\end{equation*} 
\begin{Example}{}{}
\label{ex:m-definition}
   \small{ 
%   [[[THIS EXAMPLE MUST BE CHANGED ACCORDINGLY]]]
%   To mitigate the risk of threats in Example 1, we identify a set of controls
%   $\mathcal{C} =$ \{\emph{$c_1$ (Ensuring data minimization), $c_2$ (Enabling data deletion), $c_3$ (Ensuring secure storage), $c_4$ (Logging access to personal data), $c_5$ (Ensuring data anonymization)}\}\ where $c_1$ and $c_2$ are devoted to mitigate $T_1$, $c_3$ and $c_4$ to $T_2$, and $c_5$ to $T_3$.
% We define $\mu_{T}:\mathcal{C}_T\to [0..1)$ where ${T\in \mathcal{T}}$ and $\mathcal{T}$ is defined as in Example 1, as follows:
For simplicity, we consider three possible values in the co-domain of $\mu_{T_1}$, $\mu_{T_2}$, and $\mu_{T_3}$, namely $0$ (the control does not mitigate the threat), $0.5$ (the control partially mitigates the threat), and $1$ (the control eliminates the threat).  
%as the level of threat mitigating by a control $c$, which means it is not, partially, or fully effective in mitigating the threat $T$, respectively.
Continuing the previous examples, the mitigation mappings for $T_1$, $T_2$, and $T_3$ can be defined as follows:
% $\{\mathcal{C}_T\}_{T\in \mathcal{T}}$ are as follow: 
% $\mu_{T}:\mathcal{C}_T\to [0..1)$
% for ${T\in \mathcal{T}}$ the $\mu_{T}(c)$ is as follow:
   \begin{center}
    \scriptsize
    \begin{tabular}[t]{c|c}
         $\langle\mu_{T_1}(c_1), \mu_{T_1}(c_2)\rangle$ &$m(T_1)$       \\ \hline\hline
                                $\langle0,0\rangle$     & 0   \\
                                $\langle0,0.5\rangle$     & 0.25   \\
                                $\langle0.5,0\rangle$     & 0.25   \\
                                $\langle0.5,0.5\rangle$     & 0.5   \\
                                $\langle1,0\rangle$     & 0.5    \\
                                $\langle0,1\rangle$    & 0.5    \\
                                $\langle1,0.5\rangle$     & 0.75   \\
                                $\langle0.5,1\rangle$     & 0.75   \\
                                % $C_{T1}$      &$m_{T1}(C)$       \\ \hline\hline
                                % \{$c_1$\}     & 0.5   \\
                                % \{$c_2$\}     & 0.25    \\
                                % \{$c_1,c_2$\} & 0.75  
     \end{tabular}                          
    \begin{tabular}[t]{c|c} 
        $\langle\mu_{T_2}(c_3), \mu_{T_2}(c_4)\rangle$      &$m(T_2)$       \\ \hline\hline
                                $\langle0,0\rangle$     & 0   \\
                                $\langle0,0.5\rangle$     & 0.25   \\
                                $\langle0.5,0\rangle$     & 0.25   \\
                                $\langle0.5,0.5\rangle$     & 0.5   \\
                                $\langle1,0\rangle$     & 0.5    \\
                                $\langle0,1\rangle$    & 0.5    \\
                                $\langle1,0.5\rangle$     & 0.75   \\
                                $\langle0.5,1\rangle$     & 0.75   \\
                                % $C_{T2}$       &$m_{T2}(C)$ \\ \hline\hline
                                % \{$c_3$\}      & 0.25  \\
                                % \{$c_4$\}      & 0.75  \\
                                % \{$c_3, c_4$\} & 0.5  
    \end{tabular}
    \begin{tabular}[t]{c|c} 
                    $\langle\mu_{T_3}(c_5)\rangle$      &$m(T_3)$       \\ \hline\hline
                                $\langle0\rangle$     & 0   \\
                                $\langle0.5\rangle$     & 0.5   
                                % $C_{T3}$       &$m_{T3}(C)$\\ \hline\hline
                                % \{$c_5$\}      & 0.5  
    \end{tabular}
    \end{center}
    where the first column of each table lists all possible mitigation vectors that are assigned to the controls of $\mathcal{C}_{T_1}$, $\mathcal{C}_{T_2}$, and $\mathcal{C}_{T_3}$, respectively, when considering an arbitrary total order on the controls (in our case $c_i$ comes before $c_j$ if $i<j$ for $i,j\in \{1, ..., 5\}$.
    % the minimal elements in $\mathbb{P}(\mathcal{C})$ for which the value of the mapping $m_T$ is non-zero, and the second provides the assigned mitigation levels for each of these subsets.
    For instance, the vector $\langle0.5,0\rangle$ means that $c_1$ partially mitigates $T_1$ whereas $c_2$ has no mitigation effect on $T_1$. % in the first table represents the mitigation levels for the specified controls' $T_1$, where $c_1$ is $0.5$ (partially) and $c_2$ is $0$ (not implemented).
    % the assigned mitigation level for subset (\{$c_1,c_2$\}) to mitigate $T_1$ is $0.75$.
    % The impact residue $\mathit{ir}_s(T,C)= i_s(T)\cdot (1-m_T(C))$ for each of the subsets considered in the tables above is computed as follows: 
    }
%     \begin{alignat*}{2}\scriptsize
%     \begin{aligned} & \begin{cases}
%             {ir}_{s_{1}}(T_1, \{c_1\})= 0.3 & \\
%             {ir}_{s_{1}}(T_1, \{c_2\})= 0.525 & \\
%             {ir}_{s_{1}}(T_1, \{c_1, c_2\})= 0.175 & \\
%             --------- \\
%             {ir}_{s_{1}}(T_2, \{c_3\})= 0.15 & \\
%             {ir}_{s_{1}}(T_2, \{c_4\})= 0.05 & \\
%             {ir}_{s_{1}}(T_2, \{c_3, c_4\})= 0.1 & \\
%             --------- \\
%             % {ir}_{s_{1}}(T_3, \emptyset)= 0.3 & \\
%             {ir}_{s_{1}}(T_3, \{c_5\})= 0.15 & \\
%       \end{cases}\\
%       \end{aligned}
%       \begin{aligned}
%       & \begin{cases}
%             {ir}_{s_{2}}(T_1, \{c_1\})= 0.15 & \\
%             {ir}_{s_{2}}(T_1, \{c_2\})= 0.225 & \\
%             {ir}_{s_{2}}(T_1, \{c_1, c_2\})= 0.075 & \\
%             --------- \\
%             {ir}_{s_{2}}(T_2, \{c_3\})= 0.375 & \\
%             {ir}_{s_{2}}(T_2, \{c_4\})= 0.125 & \\
%             {ir}_{s_{2}}(T_2, \{c_3, c_4\})= 0.25 & \\
%             --------- \\
%             % {ir}_{s_{2}}(T_3, \emptyset)= 0.6 & \\
%             {ir}_{s_{2}}(T_3, \{c_5\})= 0.3 & \\
%       \end{cases} \\
%       \end{aligned}
%     \end{alignat*}
   
%   \small{where each value above is derived from the control subsets and the impact levels for each threat.
%     }
 \end{Example}

\begin{Example}{}{}
\label{ex:irs-definition}
\small{
  From the definitions of $i_s(T)$ and $m(T)$ in Examples 1 and 3, respectively, we can compute the impact residue $\mathit{ir}_s(T)= i_s(T)\cdot (1- m(T))$ for each mitigation vector in Example 3 as follows: 
    }
    \begin{center}
    \scriptsize
    \begin{tabular}{c|c|c}{\centering}
    % $\mathcal{T}$ &$i_{S_1}$&$i_{S_2}$ \\ \hline\hline
    % $T_1$ & 0.6       & 0.2          \\ 
    % $T_2$ & 0.3       & 0.5         \\
    $T$  &${ir}_{s_{1}}(T)$      &${ir}_{s_{2}}(T)$ \\ \hline\hline
    \multirow{8}{*}{$T_1$}& $0.6\times(1-0)=0.6$ & $0.3\times(1-0)=0.3$  \\
                     & $0.6\times(1-0.25)=0.45$ &$0.3\times(1-0.25)=0.225$ \\
                     & $0.6\times(1-0.25)=0.45$ & $0.3\times(1-0.25)=0.225$ \\
                     & $0.6\times(1-0.5)=0.3$ & $0.3\times(1-0.5)=0.15$ \\
                     & $0.6\times(1-0.5)=0.3$ &$0.3\times(1-0.5)=0.15$  \\
                     & $0.6\times(1-0.5)=0.3$  &$0.3\times(1-0.5)=0.15$ \\
                     & $0.6\times(1-0.75)=0.15$ &$0.3\times(1-0.75)=0.06$ \\
                     & $0.6\times(1-0.75)=0.15$ &$0.3\times(1-0.75)=0.06$  \\\hline
    \multirow{8}{*}{$T_2$}& $0.2\times(1-0)=0.2$ & $0.5\times(1-0)=0.5$\\
                     & $0.2\times(1-0.25)=0.15$&$0.5\times(1-0.25)=0.375$ \\
                     & $0.2\times(1-0.25)=0.15$ &$0.5\times(1-0.25)=0.375$\\
                     & $0.2\times(1-0.5)=0.1$&$0.5\times(1-0.5)=0.25$   \\
                     & $0.2\times(1-0.5)=0.1$&$0.5\times(1-0.5)=0.25$   \\
                     & $0.2\times(1-0.5)=0.1$ &$0.5\times(1-0.5)=0.25$ \\
                     & $0.2\times(1-0.75)=0.05$&$0.5\times(1-0.75)=0.125$\\
                     & $0.2\times(1-0.75)=0.05$&$0.5\times(1-0.75)=0.125$ \\\hline
    \multirow{2}{*}{$T_3$}& $0.3\times(1-0)=0.3$ & $0.6\times(1-0)=0.6$ \\
                     & $0.3\times(1-0.5)=0.15$ & $0.6\times(1-0.5)=0.3$ \\
    \end{tabular}
    \end{center}
    
    % \begin{alignat*}{2}\scriptsize
    % \begin{aligned} & \begin{cases}
    %         {ir}_{s_{1}}(T_1)=i_s(T_1).(1-m(T_1)= 0.3 & \\
    %         {ir}_{s_{1}}(T_1, <0.5,0>)= 0.525 & \\
    %         {ir}_{s_{1}}(T_1, <0,0.5>)= 0.175 & \\
    %         {ir}_{s_{1}}(T_1, <0.5,0.5>)= 0.175 & \\
    %         {ir}_{s_{1}}(T_1, <1,0>)= 0.175 & \\
    %         {ir}_{s_{1}}(T_1, <0,1>)= 0.175 & \\
    %         {ir}_{s_{1}}(T_1, <1,0.5>)= 0.175 & \\
    %         {ir}_{s_{1}}(T_1, <0.5,1>)= 0.175 & \\
    %         --------- \\
    %         {ir}_{s_{1}}(T_2, \{c_3\})= 0.15 & \\
    %         {ir}_{s_{1}}(T_2, \{c_4\})= 0.05 & \\
    %         {ir}_{s_{1}}(T_2, \{c_3, c_4\})= 0.1 & \\
    %         --------- \\
    %         % {ir}_{s_{1}}(T_3, \emptyset)= 0.3 & \\
    %         {ir}_{s_{1}}(T_3, \{c_5\})= 0.15 & \\
    %   \end{cases}\\
    %   \end{aligned}
    %   \begin{aligned}
    %   & \begin{cases}
    %         {ir}_{s_{2}}(T_1, \{c_1\})= 0.15 & \\
    %         {ir}_{s_{2}}(T_1, \{c_2\})= 0.225 & \\
    %         {ir}_{s_{2}}(T_1, \{c_1, c_2\})= 0.075 & \\
    %         --------- \\
    %         {ir}_{s_{2}}(T_2, \{c_3\})= 0.375 & \\
    %         {ir}_{s_{2}}(T_2, \{c_4\})= 0.125 & \\
    %         {ir}_{s_{2}}(T_2, \{c_3, c_4\})= 0.25 & \\
    %         --------- \\
    %         % {ir}_{s_{2}}(T_3, \emptyset)= 0.6 & \\
    %         {ir}_{s_{2}}(T_3, \{c_5\})= 0.3 & \\
    %   \end{cases} \\
    %   \end{aligned}
    % \end{alignat*}
   
  \small{
  where the second and third columns represent the computed impact residues under all possible mitigation mappings and the corresponding threat (in the rows) for $s_1$ and $s_2$, respectively.
  For instance, the impact residue under the mitigation vector $\langle 0.5,1 \rangle$ for $T_1$ from the point of view of $s_1$ is $0.15$ whereas it is $0.06$ for $s_2$.
  Recalling that $\mathit{oir}(s)= \Sigma_{T\in \mathcal{T}} \mathit{ir}_s(T)$, the overall impact residue for $s_1$ is $\mathit{oir}(s_1)= 0.6+0.2+0.3= 1.1$ and for $s_2$ is $\mathit{oir}(s_2)= 0.3+0.5+0.6= 1.4$ where $\langle\mu_{T_1}(c_1), \mu_{T1}(c_2)\rangle = \langle0,0\rangle$, $\langle\mu_{T_2}(c_3),  \mu_{T_2}(c_4)\rangle= \langle0,0\rangle$, and $\langle\mu_{T_3}(c_5)\rangle= \langle0\rangle$. 
  
%   when no controls have been implemented for each threat (i.e., the first inner rows)
    }
 \end{Example}

%Under the (simplifying) assumption that threats in $\mathcal{T}$ are independent, 
The \emph{Multi-Stakeholder Risk Minimization Problem} (MSRMP) amounts to solve the following multi-objective optimization problem:
\begin{equation}
    \label{eq:msrmp}
    \begin{array}{l}
    \mathit{min}_{\langle \mu_T \rangle_{T\in \mathcal{T}}} ~~ 
        \langle \mathit{oir}(s) \rangle_{s\in \mathcal{S}}
        \end{array}
\end{equation}
% \begin{equation}
%     \label{eq:msrmp}
%     \mathit{min}_{C\in \mathbb{P}(\mathcal{C})} ~~ 
%         \langle \frac{1}{|\mathcal{T}|} \Sigma_{T\in \mathcal{T}}
%         % \Pi_{T\in \mathcal{T}}
%         ~ i_s(T)*(1-m_T(C))\rangle_{s\in \mathcal{S}}
% \end{equation}
% where $\mathit{ir}_s(T,C)= i_s(T)\cdot (1-m_T(C))$ is the impact residue of the threat $T$ after the adoption of the controls in $C$, $\mathit{ir}_s^{\mathit{avg}}(C)=\frac{1}{|\mathcal{T}|} \Sigma_{T\in \mathcal{T}}~ \mathit{ir}_s(T,C)$ is the overall impact residue for the stakeholder $s$, and $\langle~\rangle_{s\in \mathcal{S}}$ is the vector of all impact residues according to an arbitrary total order over $\mathcal{S}$.
where $\langle~\rangle_{T\in \mathcal{T}}$ and $\langle~\rangle_{s\in \mathcal{S}}$ are the vectors of all mitigation mappings and overall impact residues (under the associated mitigation mappings) according to arbitrary total orders over $\mathcal{T}$ and $\mathcal{S}$, respectively.  In other words, the MSRMP consists of finding the vector of mitigation mappings that allows for minimizing the overall impact residues of the stakeholders. 
% Minimizing all impact residues may turn out to be an impossible task as the following example shows.  Consider two stakeholders and the following two vectors of impact residues: $\langle 1,2\rangle$ and $\langle 2,1\rangle$; obviously, the first stakeholder prefers the former whereas the second the latter.  To precisely define which are the solutions to~\eqref{eq:msrmp}, we need to use the notion of Pareto optimality that in our case can be stated as follows.  
%According to Section~\ref{subsec:moop}, such mitigation mappings correspond to the Pareto optimal solutions of (\ref{eq:msrmp}).
A solution of (\ref{eq:msrmp}) is a vector $\langle \mu_T \rangle_{T\in \mathcal{T}}$ of mitigation mappings that is Pareto optimal (see, e.g.,~\cite{marler2004survey}), i.e.\ it is such that if there does not exist another vector $\langle \mu'_T \rangle_{T\in \mathcal{T}}$ of mitigation mappings such that $\mathit{oir}(s)\leq \mathit{oir}'(\overline{s})$ for each $s\in \mathcal{S}$ and $\mathit{oir}'({\overline{s}})< \mathit{oir}({s})$ for at least one $\overline{s}\in \mathcal{S}$ where $\mathit{oir}$ and $\mathit{oir}'$ are the overall impact residues under the family $\{ \mu_T\}_{T\in \mathcal{T}}$ and $\{ \mu'_T\}_{T\in \mathcal{T}}$ of mitigation mappings, respectively.

We make two observations.  First, \eqref{eq:msrmp} considers only the impact and not the likelihood since, as already discussed earlier, we assume that the stakeholders in $\mathcal{S}$ agree on both the set $\mathcal{T}$ of threats and their likelihood.  As a consequence, minimizing the impact is equivalent to minimizing the risk since the latter is the product of impact and likelihood, and it is a constant and positive value for each stakeholder in $\mathcal{S}$.  This is a natural assumption to make in the context of the GDPR whereby the data controller is accountable for the risk assessment and needs to guarantee that the risks of the data subject are kept to a minimum.  The second observation is about solving~\eqref{eq:msrmp}.  Indeed, it is possible to re-use the cornucopia of techniques available for Multi Objective Optimization Problem (MOOP); see, e.g.,~\cite{marler2004survey}.  However, for some of the techniques to be applicable, it is crucial to have a definition of the functions $i_s$ and $\mu_T$ for $T\in \mathcal{T}$ in closed form.  This is rarely the case for the use case scenarios we have in mind.  Instead experts are typically able to define both $i_s$ and $\mu_T$ as discrete functions, i.e.\ by associating a given impact level with a certain threat for $i_s$ and quantifying the amplitude of the mitigation associated to a given control in  $\mathcal{C}_T$ for $\mu_T$. 
The examples above present this kind of definitions for such functions by using tables.
%[[[SAY SOMETHING ABOUT THE DIFFICULTY OF SOLVING THE PROBLEM THAT DEPENDS ON HOW THE FUNCTIONS $i_S$ and $m_T$ ARE DEFINED... WE NEED TO ARGUE THAT WHILE IT IS REASONABLE TO EXPECT THAT EXPERTS OF EACH STAKEHOLDER ARE ABLE TO ASSOCIATE AN IMPACT LEVEL WITH A THREAT, I.E.\ WE CAN ASSUME THAT THEY ARE ABLE TO PROVIDE A SPECIFICATION OF THE FUNCTION $i_s$ IN TABULAR FORMAT, I.E.\ THEY EXPLICITLY ASSOCIATE A IMPACT LEVEL WITH A GIVEN THREAT, THE SAME CANNOT BE SAID FOR THE FUNCTION $m_T$ AS IT REQUIRES TO UNDERSTAND THE COMBINED EFFECT OF ALL THE AVAILABLE CONTROLS ON EACH THREAT... FOR THIS REASON, WE SIMPLIFY THE TASK OF EXPERTS AND ASK THEM TO DEFINE A FUNCTION (CALLED $a_T$ BELOW) THAT ASSOCIATES THE EFFECT OF A CONTROL ON A SINGLE THREAT.  IF THIS IS AVAILABLE, WE GIVE A METHOD TO DEFINE $m_T$ OUT OF the $a_T$s, I.E.\ HOW THE  COMBINED EFFECTS OF THE VARIOUS CONTROLS AFFECT A THREAT... THIS SHOULD BE EXPLAINED INSTEAD OF THE TOO A SHORT EXPLANATION BELOW... PROBABLY WE NEED TO COMBINE THIS WITH THE TEXT BELOW...]]]

As a consequence of the two observations above, we make the following assumptions.  First, each stakeholder $s$ in $\mathcal{S}$ provides a definition of the mapping $i_s$ as a finite set of pairs of the form $(T,\mathit{il})$ where $T$ is a threat in $\mathcal{T}$ and $\mathit{il}$ is an impact level in a finite set $\mathcal{I}$ of values (i.e., $\mathcal{I}=\{0,1,2,3,4\}$ where $0$ denotes a negligible impact, $4$ a dramatic impact, and the values in between increasing values).  Second, for each threat $T$ in $\mathcal{T}$, the stakeholder in charge of the risk management process (i.e., the data controller in the case of the GDPR) defines the mapping $\mu_T:\mathcal{C}_T\to \mathcal{A}$ with $\mathcal{A}$ a finite set of values in the interval $[0..1]$; in other words, $\mu_T$ is specified as a finite set of pairs of the form $(c,p)$ where $c$ is a control in $\mathcal{C}$ and $p$ is the amplitude of the mitigation of the impact of the threat $T$ when adopting the control $c$.  For instance, we can take $\mathcal{A}=\{0, 0.5, 1\}$, so that $\mu_T(c)=0$ means that control $c$ has no effect in mitigating the threat $T$, $\mu_T(c)=0.5$ has partial effect on $T$, and $\mu_T(c)=1$ has full effect. % A control $c$ that is intended for a certain threat $T\neq T'$ may have no or only a partial effect on the threat $T'$ and thus $a_T(c)=0$ or $a_T(c)=0.5$, respectively.\footnote{We observe that it is possible to consider more values as the co-domain of the mapping $a_T$ to a have a more fine-grained characterization of the effects of a control on a threat.  Here, we consider only three for the sake of simplicity.  The discussion that follows holds also for a larger number of values.}   A control $c$ especially designed to mitigate the threat $T$ should have full effect on it, i.e.\ $a_T(c)=1$.  Thus, 
%Then, we define the measure $m_T(C)$ of the combined effects of the controls in a set $C\subseteq \mathcal{C}_T$ on a threat $T$ in $\mathcal{T}$ as
%\begin{equation}
%\label{eq:msrmp1}
%   m_T(C) = \frac{\Sigma_{c\in C} ~ a_T(c)}{|{C}|} 
%\end{equation}
%In other words, the co-domains of the mappings $i_s$ and $\mu_T$ are finite sets of  values.    
Under these assumptions, we obtain an instance of~\eqref{eq:msrmp} that
% we are able to rewrite the MSRMP~\eqref{eq:msrmp} as follows:
% \begin{equation}
%     \label{eq:msrmp1}
%     \mathit{min}_{C\in \mathbb{P}(\mathcal{C})} ~~ 
%         \langle \frac{1}{|\mathcal{T}|} \Sigma_{T\in \mathcal{T}}~ i_s(T)\cdot (1- m_T(C))\rangle_{s\in \mathcal{S}} .
%         % \frac{\Sigma_{c\in C} ~ a_T(c)}{|{C}|}
% \end{equation}
belongs to a particular class of MOOP called Multi Objective Combinatorial Optimization Problems (MOCOPs); see, e.g.,~\cite{mocop}.  
We observe that finding all Pareto optimal solutions of such instances of~\eqref{eq:msrmp} requires, in the worst case, to search among $\Pi_{T\in \mathcal{T}}( k^{|\mathcal{C}_T|}-1)$ candidate sets of controls for $k=|\mathcal{A}|$ the number of distinct real values in the co-domain of the mappings $\mu_T$ for all $T$ in $\mathcal{T}$.  The $-1$ in the expression considers that it is never the case that all controls in $\mathcal{C}_T$ will be adopted; this is a reasonable assumption because of multiple reasons including lack of skills to manage several different technologies on which the controls are based and constraints in costs.  
% PLEASE MAJID TAKE A LOOK AT THE REST OF THIS PARAGRAPH THAT I HAVE ADDED TO CLARIFY A COUPLE OF THINGS... 
Indeed, this implies the decidability of the instances of the MSRMP that we consider in the rest of the work.  We observe that, despite their decidability, solving these instances of the MSRMP may be quite a challenge from a computational point of view because the number of possible solutions in which to search for the optimal ones is exponential in the size of $\mathcal{C}_T$ for $T\in \mathcal{T}$.  In the rest of this section, we describe a strategy to manage this problem and in Section~\ref{subsec:experimental-results}, we propose an experimental evaluation of some refinements and study the scalability of the proposed approach in practice.  

\begin{Example}{}{}
\small
As described above, by considering $k=3$ possible values for the mappings
$\mu_{T_1}$, $\mu_{T_2}$, and $\mu_{T_3}$ introduced in Example 3,
  the search for finding optimal solutions is among $\Pi_{T\in \mathcal{T}}( k^{|\mathcal{C}_T|}-1)$= $(3^{2}-1)\times(3^{2}-1)\times(3^{1}-1)=128$ candidates. 
  Note that we do not consider the situation in which all controls are in place as this would yield a risk equal to zero, thereby making the search for optimal solutions trivial.  
  This is reasonable in practice since, as already observed, it is unlikely that the stakeholders will be able to adopt all security controls in $\{\mathcal{C}_T\}_{T\in \mathcal{T}}$ because of other constraints such as those related to budget and required security skills for their deployment.
\end{Example}

To simplify the solution of the instances of~\eqref{eq:msrmp}, we consider an associated problem derived from~\eqref{eq:msrmp}, by introducing a variable $x_T$ to replace $1-m(T)$ and
% the expression $1-\frac{\Sigma_{c\in C} ~ \mu_T(c)}{|{C}|}$ to
obtain:
% can be transformed into the following problem:  
\begin{equation}
    \label{eq:msrmp2}
    \begin{array}{l}
    \mathit{min}_{\langle x_T\rangle_{T\in\mathcal{T}}} ~~ 
        \langle\frac{1}{|\mathcal{T}|} \Sigma_{T\in \mathcal{T}}~ (i_s(T)*x_T) \rangle_{s\in \mathcal{S}} \\
        \mbox{subject to } x_T\in \left\{ 1-m(T) \right\}
        \mbox{ for each } T\in \mathcal{T}
    \end{array}
\end{equation}
where $m(T)$ is the expression defined in~\ref{eq:mitigation-abbreviation}, $\langle x_T\rangle_{T\in \mathcal{T}}$ is the vector of variables
representing mitigation amplitudes 
% replacing expressions for impact residues of threats 
when considering an arbitrary total order over $\mathcal{T}$. % and $X_T$ is defined as follows:
% \begin{equation*}
 %   {X}_T = \left\{ 1-m(T) \right\}.
%\end{equation*}
% where $\mathcal{C}_{\mathcal{T}}$ contains a subset of $\mathcal{C}$ for each threat $T$ of $\mathcal{T}$.  
For each threat $T$ in $\mathcal{T}$, we have that $|\left\{ 1-m(T) \right\}|$ is the number of distinct sum values, divided by the number of controls in $\mathcal{C}_T$, that can be obtained by adding values in $\mathcal{I}$ (that, in our examples, is the set $\{0,0.5,1\}$) according to a $\mu_T$ that induces a value $m(T)$.
% associated to a control in the set $C_T$ associated to $T$.  
The space of solutions of the modified version of~\eqref{eq:msrmp2},
% i.e.\ the problem obtained from~\eqref{eq:msrmp}
% by substituting $\tilde{X}_T$ in place of $X_T$, 
is thus $\Pi_{T\in \mathcal{T}} |\{1-m(T)\}|$ which may be remarkably less than $\Pi_{T\in \mathcal{T}} (k^{|\mathcal{C}_T|}-1)$.
For instance, consider Example 3, the first two tables contain 8 different mitigation vectors with only 4 different values for the function $m(\cdot)$.

\begin{Example}{}{}
\label{ex:m-pareto}
\small
% \ref{ex:m-definition}
   Recall Example 3, consider only the values of $m(T)$ that are distinct, and derive the values $x_T= 1-m(T)$ for each $T\in \{T_1, T_2, T_3\}$: 
   \begin{center}
    \scriptsize
    \begin{tabular}[t]{c|c}
                           $m(T_1)$&$x_{T_1}$ \\ \hline\hline 
    $0$ & $1$   \\ 
    $0.25$ & $0.75$  \\ 
    $0.5$ &$0.5$         \\ 
    $0.75$ &$0.25$         \\ 
     \end{tabular}                          
    \begin{tabular}[t]{c|c} 
                        $m(T_2)$&$x_{T_2}$ \\ \hline\hline 
    $0$ & $1$   \\ 
    $0.25$ & $0.75$  \\ 
    $0.5$ &$0.5$         \\ 
    $0.75$ &$0.25$         \\ 
    \end{tabular}
    \begin{tabular}[t]{c|c} 
                    $m(T_3)$&$x_{T_3}$ \\ \hline\hline 
    $0$ & $1$   \\ 
    $0.5$ & $0.5$  \\ 
  
    \end{tabular}
    \end{center}
   The set of possible solutions of (\ref{eq:msrmp2}) is the set of all triples of the form $\langle x_{T_1}, x_{T_2}, x_{T_3} \rangle$ whose values are taken from the three tables above and thus the size of such a set is % set will reduce remarkably. For instance here, ${{X}_{T_1}}$ = ${{X}_{T_2}}$ = $\{1,0.75,0.5,0.25\}$, and $X_{T_3}$ = $\{1, 0.5 \}$, then $\Pi_{T\in \mathcal{T}} |{X}_T|$ equals 
   $4\times4\times2= 32$. Observe that this is one-fourth of the size of the set of potential solutions to the original problem (\ref{eq:msrmp}), namely 
   $\Pi_{T\in \{T_1, T_2, T_3\}}(k^{\mathcal{|C_T|}}-1)=(3^2-1)\cdot(3^2-1)\cdot (3^1-1)=128$. 
   %The solution set $X_T$ contains 32 candidates plotted in the following figure. We use Pareto optimality to pick solutions.
   For larger problem instances, the reduction is much more substantial as we will see in Section\ref{subsec:experimental-results} below.
   By considering the $32$ triples $\langle x_{T_1}, x_{T_2}, x_{T_3} \rangle$, we can derive the values of the overall impact values for the two stakeholders by recalling that 
   $\mathit{oir}(s) =  \mathit{ir}_s(T_1) + \mathit{ir}_s(T_2) + \mathit{ir}_s(T_3)$, $\mathit{ir}_s(T) = i_s(T)\cdot (1-m(T))$ from (\ref{eq:impact}), (\ref{eq:mitigation-abbreviation}, and $x_T = 1-m(T)$ for $s\in \{s_1, s_2\}$ and for $T\in \{T_1, T_2, T_3\}$.  Also, recall that the definition of $i_s(\cdot)$ can be found in Example 4. 
%   \ref{ex:irs-definition}
   The pairs $(\mathit{oir}(s_1), \mathit{oir}(s_2))$ so computed are plotted in Figure \ref{fig:solution} where the x-axis shows the values of $\mathit{oir}(s_1)$  and the y-axis those of $\mathit{oir}(s_2)$.  
   \begin{center}
     \centering
    %  \caption{Solutions}
		\includegraphics[width=0.7\textwidth, height=4.6cm]{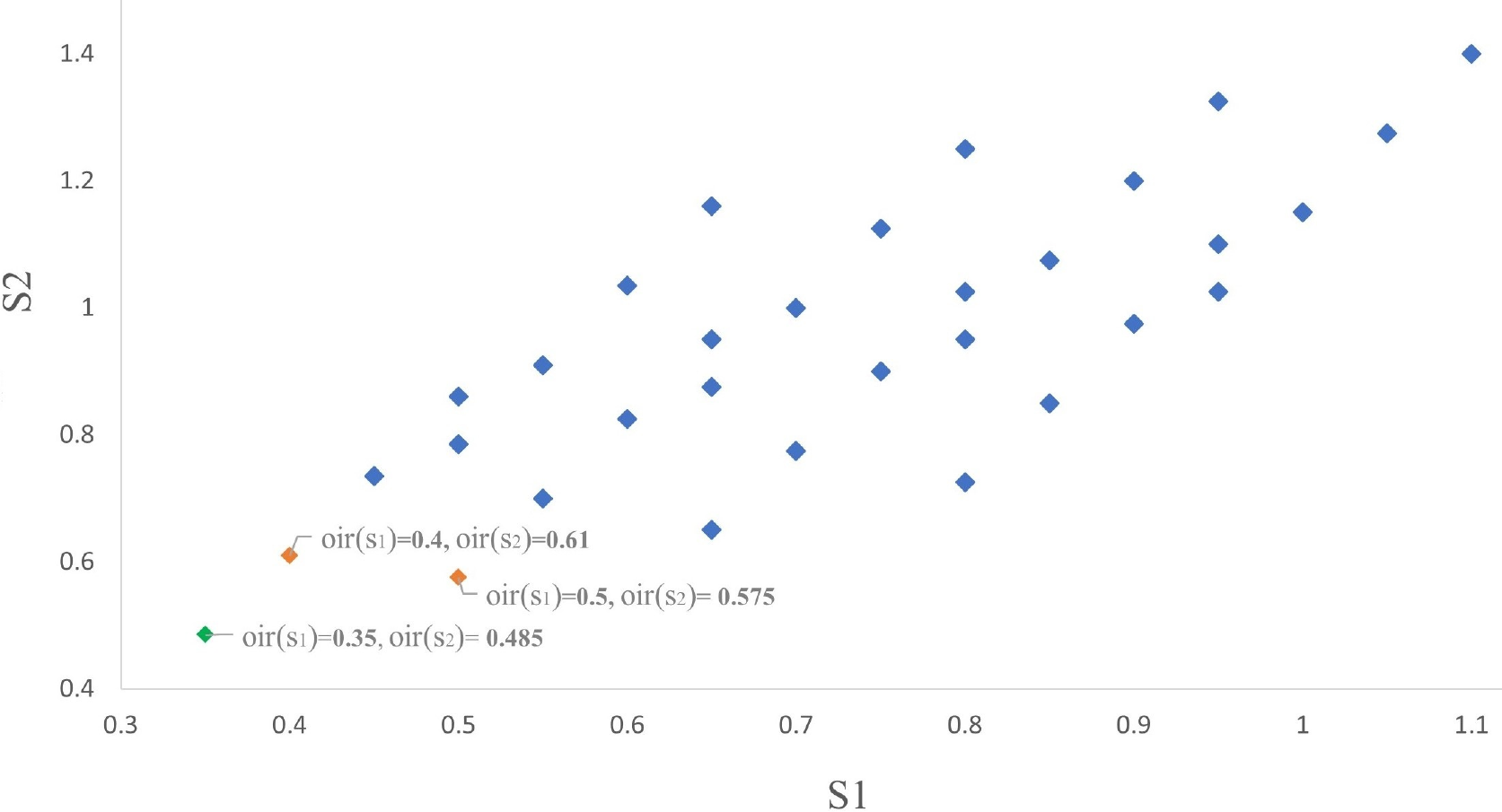}
% 		\caption{ss} 
         \captionof{figure}{The solution points.}
		\label{fig:solution}
  \end{center}
  It is then immediate to see that the point $(0.35,0.485)$ at the bottom left (in green) is the Pareto optimal solution. We also observe that the two points in orange are not dominated by any other points but the optimal one.  
\end{Example}

% \begin{Example}{}{}
% \small
%  Continuing Example 5, the solution set $X_T$ contains 32 candidates plotted in the following figure. We use Pareto optimality to pick solutions. 
%   \begin{center}
%      \centering
% 		\includegraphics[width=\textwidth]{Figures/Solutions.jpg}
% % 		\caption{ss} 
%         % \captionof{figure}{The solution points}
% 		\label{fig:solution}
%   \end{center}
%  In this figure, the $\mathit{oir}(s_1)$ and $\mathit{oir}(s_2)$ are represented in the x-axis and y-axis, respectively.
%  The point at the bottom left (the green color) is the optimal solution where $\mathit{oir}(s_1)=0.35$ and $\mathit{oir}(s_2)=0.485$ and it dominates all other solution points in this figure.
%  In this figure, the other two solutions shown in orange, are not dominated by any other points except the optimal one.
% \end{Example}

Indeed, it is possible to find solutions of~\eqref{eq:msrmp} corresponding to those of the simplified version of~\eqref{eq:msrmp2} by adapting the procedure above.  Let $\langle x_T^*\rangle_{T\in \mathcal{T}}$ be a solution for~\eqref{eq:msrmp2}.  By definition and the simplifying assumption above, there must exist $\mu_T^*$ such that $x_T^*= 1-m(T) = 1-\frac{\Sigma_{c\in \mathcal{C}_T} ~ \mu_T^*(c)}{|\mathcal{C}_T|}$ for each $T\in \mathcal{T}$ and it is thus immediate to discover all the solutions of~\eqref{eq:msrmp}.  
% By observing that $C^*\subseteq C_T$ (instead of $C^*\subseteq \mathcal{C}$ as it was the case above), it is also easy to understand the speed-up obtained by the introduction of $\mathcal{C}_{\mathcal{T}}$ in deriving the solutions of~\eqref{eq:msrmp} from those of~\eqref{eq:msrmp2} and not only for solving the latter. 

\begin{Example}{}{}
\label{ex:mapping-back}
\small
We explain how it is possible to derive the sets of controls associated to a certain triple $\langle x_{T_1}^*, x_{T_2}^*, x_{T_3}^* \rangle$.  To illustrate, we consider the (orange) point in Figure \ref{fig:solution} with coordinates $(0.4,0.61)$ that is associated to the triple $\langle x_{T_1}^*, x_{T_2}^*, x_{T_3}^* \rangle = \langle 0.25, 0.5, 0.5\rangle$.
%First, the impact residue $\mathit{ir}_s(T)=i_s(T)\cdot (1-m(T))$ for each threat $T\in \{T_1, T_2, T_3\}$ and stakeholder $s\in  \{s_1, s_2\}$ is derived (just recall (\ref{eq:impact}) and (\ref{eq:mitigation-abbreviation})): 
   %\begin{center}
%   \scriptsize
    %\begin{tabular}{c|c|c}{\centering}
    %$T$  &${ir}_{s_{1}}(T)$      &${ir}_{s_{2}}(T)$ \\ \hline\hline
    %$T_1$ & 0.15       & 0.06         \\\hline 
    %$T_2$& 0.1       & 0.25         \\\hline 
    %$T_3$ & 0.15       & 0.3          \\ 
    %\end{tabular}
    %\end{center}
%From the table in Example 4, we are able to retrieve $m(T)$ values. For instance, $\mathit{oir}(s_1)={ir}_{s_{1}}(T_1)$+${ir}_{s_{1}}(T_2)$+${ir}_{s_{1}}(T_3)=0.4$. To retrieve $m(T_1)$, we know 
%$\mathit{ir}_{s_{1}}(T_1)= i_s(T_1)\cdot (1- m(T_1))$ which means $m(T_1)= 1-(\frac{\mathit{ir}_{s_{1}}(T_1)}{ i_s(T_1)})=1-(\frac{0.15}{0.6})=1-0.25=0.75$. 
%In the same way, we can obtain $m(T_2)= m(T_3)= 0.5$ which means $x_{T_1}^*= 0.25$, and $x_{T_2}^*= x_{T_3}^*= 0.5$.
% by looking at this table and with knowledge of Example 4, we understand $m(T)$ values are $m(T_1)= 0.75$, and $m(T_2)= m(T_3)= 0.5$ which means $x_{T_1}^*= 0.25$, and $x_{T_2}^*= x_{T_3}^*= 0.5$. 
%Therefore, we look under which mitigation vectors we meet again the same values for $x_{T_1}^*, x_{T_2}^*$, and $x_{T_3}^*$. 
From (\ref{eq:mitigation-abbreviation}) and $x_T = 1-m(T)$, it is immediate to derive that
$$
   \frac{\mu_{T_1}(c_1) + \mu_{T_1}(c_2)}{2} = 1 - x_{T_1}^*  \quad
   \frac{\mu_{T_2}(c_3) + \mu_{T_2}(c_4)}{2} = 1 - x_{T_2}^*  \quad
   \frac{\mu_{T_3}(c_5)}{1} = 1 - x_{T_3}^* 
$$ 
so that we are left with the problem of enumerating all mitigation mappings $\mu_{T_1}(\cdot), \mu_{T_2}(\cdot), \mu_{T_3}(\cdot)$ satisfying the three equalities above.
The following table lists all possible such mappings: 
\begin{center}
\begin{tabular}{l|cc|cc|c}
\multirow{2}{*}{} & \multicolumn{2}{l|}{$x_{T_1}^*= 0.25$} & \multicolumn{2}{l|}{$x_{T_2}^*= 0.5$} & \multicolumn{1}{l}{$x_{T_3}^*= 0.5$} \\\cline{2-6} 
               & $\mu_{T_1}(c_1)$ & $\mu_{T_1}(c_2)$ & $\mu_{T_2}(c_3)$ & $\mu_{T_2}(c_4)$ & $\mu_{T_3}(c_5)$ \\ \hline
$\mathbb{S}_1$ & 1     & 0.5   & 0.5   & 0.5   & 0.5   \\
$\mathbb{S}_2$ & 0.5   & 1     & 0.5   & 0.5   & 0.5   \\
$\mathbb{S}_3$ & 1     & 0.5   & 1     & 0     & 0.5   \\
$\mathbb{S}_4$ & 0.5   & 1     & 1     & 0     & 0.5   \\
$\mathbb{S}_5$ & 1     & 0.5   & 0     & 1     & 0.5   \\
$\mathbb{S}_6$ & 0.5   & 1     & 0     & 1     & 0.5  
\end{tabular}
\end{center}
%This table reports six possible mitigation vector combinations where the overall risk will be $\mathit{oir}(s_1)= 0.4$ and $\mathit{oir}(s_2)= 0.61$.
%For instance, $\mathbb{S}_1$ represents $\langle\mu_{T_1}(c_1), \mu_{T_1}(c_2)\rangle$= $\langle 1, 0.5\rangle$, $\langle\mu_{T_2}(c_3), \mu_{T_2}(c_4)\rangle$= $\langle 0.5, 0.5\rangle$ and $\langle\mu_{T_3}(c_5)\rangle$= $\langle 0.5\rangle$.
% the impact residue result for all these solutions would be same.
\end{Example}
The obvious question is the computational complexity of enumerating all possible mitigation mappings $\mu_T(\cdot)$ such that 
\begin{equation}
\label{eq:mapping-back-eqs}
    \frac{\Sigma_{c\in \mathcal{C}_T} \mu_T(c)}{|\mathcal{C}_T|}) = 1 - x_T^*
\end{equation}
for each $T\in \mathcal{T}$; notice that the three equalities in Example~7 are instances of (\ref{eq:mapping-back-eqs}).  Indeed, if there exists a (practically) efficient algorithm to enumerate the mitigation mappings satisfying (\ref{eq:mapping-back-eqs}), we can hope that solving instances of (\ref{eq:msrmp2}) and then using such an algorithm to derive the corresponding solutions of (\ref{eq:msrmp}) is an efficient alternative to solving directly the latter as the number of the possible solutions of (\ref{eq:msrmp2}) is smaller (as we have seen in Example 6 and even substantially so as we will see in Section \ref{subsec:experimental-results}) than those of (\ref{eq:msrmp}).  

To answer this question, we consider the Subset Sum Problem (SSP) with multiplicities \cite{ssp}, i.e.\ given a multiset $X$ of integers and an integer $s$, does any non-empty multisubset of $X$ sum to $s$?  Solving the instances of (\ref{eq:mapping-back-eqs}) for each $T\in \mathcal{T}$ is equivalent to solving an instance of the SSP under the natural assumption that $x_T^*$ and the values in $\mathcal{A}$ are real numbers that can be represented as $v\cdot 10^{-d}$ for $v$ and $d$ positive integers such that $0<v\cdot 10^{-d}<1$.  To see this, observe that all the values in $\mathcal{A}\cup \{1-x_T^*\}$ can be transformed to integers by multiplying each one by their maximum exponent $d$ when represented as $v\cdot 10^{-d}$, the integers so obtained from the values in $\mathcal{A}$ are added to the multiset $X$, each one with multiplicity equal to the number of controls in $\mathcal{C}_T$ for $T\in \mathcal{T}$, and the integer obtained from $1-x_T^*$ is set to $s$. 
Several different algorithms are available to solve this problem with different complexities ranging from exponential to (pseudo-)polynomial (see, e.g., \cite{ssp}).  The most naive algorithm (with exponential worst-case complexity) amounts to cycling through all multisubsets of $X$ and, for each one, check if it sums to $s$. 
To solve the SSP, it is possible to stop as soon as one solution is found, but in our case, we need to find all possible solutions.  
Indeed, the naive algorithm can be trivially adapted to do this, resulting in exponential best-case and worst-case complexity. 
Despite being in such a complexity class, the naive algorithm turns out to give satisfactory results in practice because the instances derived from (\ref{eq:mapping-back-eqs}) are typically small because the cardinality of $\mathcal{C}_T$ is relatively small for each $T\in \mathcal{T}$ or can be reduced by exploiting the knowledge of security experts.  
We will discuss this issue in Section \ref{sec:impl-exp} below.

\section{Defining Instances of the MSRMP }
\label{sec:new-methodology}

Our main goal is to assist in the identification of the best possible set of controls to minimize the risk for all stakeholders.  This has been formalized as solving an appropriate instance of the MSRMP introduced in Section~\ref{subsec:problem}.  To specify instances of the MSRMP in either statement (\ref{eq:msrmp}) or (\ref{eq:msrmp2}), we consider additional information that is typically available in many methodologies for risk assessment.  In the rest of this section, we first (Section~\ref{subsec:il-1st-attempt}) consider the problem statement (\ref{eq:msrmp2}) and discuss an approach to derive the risk residue $i_s$ for each stakeholder $s$ that yields a problem with a reduced search space whose solutions can be used to derive optimal mitigation mappings as explained at the end of Section~~\ref{subsec:problem}.  We will see that this approach requires the stakeholder $s$ to take several decisions that are highly subjective and this may lead to bias.  Then (Section~\ref{subsec:il-refinement}), we propose an approach that aims to reduce the level of subjectivity in defining the risk residue $i_s$ that requires to consider the general problem statement (\ref{eq:msrmp2}).  We will discuss how also in this case it is possible to first solve a problem with a reduced search space and then to derive optimal mitigation mappings.  
Both approaches require to identify a set $\mathcal{S}$ of stakeholders, a set $\mathcal{T}$ of threats, a family $\{\mathcal{C_T}\}_{T \in \mathcal{T}}$ of sets of controls (each one associated to a threat ${T \in \mathcal{T}}$), and be able to define the mapping $i_s$ that quantifies the impact level for each stakeholder $s\in \mathcal{S}$ and the residual risk $x_T$ for each threat $T$ that results from applying a certain set of controls (or, equivalently, from selecting a certain mitigation mapping $\mu_T$).  The approaches presented in Sections~\ref{subsec:il-1st-attempt} and~\ref{subsec:il-refinement} differ in the definition of $i_s$.  For this reason, we preliminary consider the definitions of the other parameters, namely $\mathcal{T}$, $\{\mathcal{C_T}\}_{T \in \mathcal{T}}$, and $x_T$.

%Our objective is to solve instances of the MSRToAP, i.e.:
%\begin{equation*}
%    \label{eq:ms1}
%    \begin{array}{l}
%    \mathit{min}_{\langle x_T\rangle_{T\in\mathcal{T}}} ~~ 
%        \langle\Sigma_{T\in \mathcal{T}}~ (i_s(T)*x_T) \rangle_{s\in \mathcal{S}} \\
%        \mbox{subject to } x_T\in \{1-m(T)\}
%    \end{array}g
%\end{equation*}
%where $m(T)=\frac{\Sigma_{c\in \mathcal{C}_T} \mu_T(c)}{|\mathcal{C}_T|}$.
%For this we need to define both $x_T$ and $i_s(T)$ for each $T \in \mathcal{T}$ and $s\in \mathcal{S}$.

% Section~\ref{subsec:data-protection}

As reviewed earlier, the literature lists several approaches (e.g.,~\cite{shostack2014threat, wuyts2015linddun})
dealing with threat identification together with appropriate mitigation controls that allow us to define the set $\mathcal{T}$ of threats and the family $\{\mathcal{C}_T\}_{T \in \mathcal{T}}$ of sets of controls associated to the threats in $\mathcal{T}$. The decision to select a method or another depends on the specific needs and specific concerns (see, e.g., the discussion in~\cite{shevchenko2018threat}). 
For instance, Microsoft STRIDE~\cite{shostack2014threat} is a well-established threat modeling to identify security threats according to a predefined classification of threat types. 
It is an acronym for \emph{Spoofing, Tampering, Repudiation, Information Disclosure, Denial of Service, }and \emph{Elevation of privilege}. 
These threat types represent the violation of the primary security properties: authentication, integrity, non-repudiation, confidentiality, availability, and authorization. 
LINDDUN~\cite{wuyts2015linddun} is another well-known threat modeling approach to identify privacy threats, and it is an acronym for \emph{Linkability, Identifiability, Non-repudiation, Unawareness, Detectability, Disclosure of information,} and \emph{Non-compliance}. Similar to STRIDE, also, these represent violations of properties characterizing different dimensions of privacy.  
For concreteness, an instance of the set $\mathcal{T}$ is shown in Table~\ref{tab:threat-table} and an instance of the family $\{\mathcal{C_T}\}_{T\in \mathcal{T}}$ can be found in the first two columns of Table~\ref{tab:threat-control} (for instance, consider $T_4=\mbox{Denial of service}$,  $\mathcal{C}_{T_4}$ is associated with three controls, namely \emph{ Enabling off-line authentication}, \emph{Network monitoring}, and \emph{Prevention mechanisms for DoS attacks like firewalls, etc.}); both are related to the running example introduced in Section~\ref{subsec:scenario}.  For the applicability of the method proposed in this work, any methodology that allows for the definition of $\mathcal{T}$ and $\{\mathcal{C_T}\}_{T\in \mathcal{T}}$ can be used. 
\begin{table}[!b] \center
    \caption{An example of possible threat scenarios and associated malicious activities in the ACME scenario}
    % \scriptsize
    \footnotesize
\begin{tabular}{l|l}
\multicolumn{1}{c|}{Threats $(\mathcal{T})$} &
  \multicolumn{1}{c}{Possible malicious activity} \\ \hline\hline
$T_1$- Unlimited data storage &
  \begin{tabular}[c]{@{}l@{}}Personal data is kept stored longer than necessary for \\ the purposes by ACME.\end{tabular} \\ \hline
\begin{tabular}[c]{@{}l@{}}
$T_2$- Unauthorized access \\ and disclosure\end{tabular} &
  \begin{tabular}[c]{@{}l@{}}Due to over-privileged or inadequate controls, insiders\\ (i.e., a medical practitioner or an ACME’s staff) modify\\ patients’ data or disclose by mistake.\end{tabular} \\ \hline
$T_3$- Linkage attack &
  \begin{tabular}[c]{@{}l@{}}Patients and their personal data can re-identify in \\ de-identified data sets by outsiders’ malicious.\end{tabular} \\ \hline
$T_4$- Denial of service &
  \begin{tabular}[c]{@{}l@{}}Attackers can disrupt the communication channel \\ between patients and the healthcare service provider \\ to prevent data from being uploaded to the server.\end{tabular} \\ \hline
$T_5$- Threat to intervenability &
  \begin{tabular}[c]{@{}l@{}}ACME does not implement a procedure (technical and\\ /or processes) that allows the patients to rectify, erase,\\  or block individual data.\end{tabular}
\end{tabular}
    \label{tab:threat-table}
\end{table}
%Concerning the definition of the set $\mathcal{G}$ of protection goals, the situation is similar as any methodology for their identification can be used; for the sake of concreteness, we define $\mathcal{G}$ to contain the goals of Section~\ref{subsec:data-protection} as they are representative of a large number of security and privacy requirements including those mentioned in the GDPR and are also relevant for the running example.  
%For the sake of simplicity and illustration, Table~\ref{tab:threat-table} shows five possible threats relevant to the running example of~\ref{subsec:scenario}.
We are left with the problem of defining $i_s$ for $s\in \mathcal{S}$ and $x_T$ for $T\in \mathcal{T}$.  Concerning the latter, recall that
%
%\subsection{\textbf{Defining Risk Residues}} % mitigation mappings}}
%
%Security and privacy controls are the administrative, technical, and physical safeguards or countermeasures to avoid, detect, counteract, or mitigate security and privacy risks within an organization~\cite{ref_35,force2018risk}.
%To establish and extract the required controls inside a system, there are several frameworks, guidelines, and best practices.
%For the sake of simplicity, we identified and mapped some potential mitigation controls for each $T \in \mathcal{T}$. The mappings between threats in $\mathcal{T}$ and controls in $\{\mathcal{C_T}\}_{T \in \mathcal{T}}$ are presented in Table\ref{tab:threat-control}, where the threats and their associated controls are reported in the first and second column, respectively. 
%For instance, for the \emph{Denial of Service} threat ($T_4$), $\mathcal{C}_{T_4}$ is associated with three controls which are (i)\emph{ Enabling off-line authentication}, (ii) \emph{Network monitoring}, and (iii) \emph{Prevention mechanisms for DoS attacks like firewalls, IDS, etc}.
%
\begin{table}[!b] \center
	\caption{Threats with associated security controls (first two columns) together with a mitigation mapping (third column) and the resulting risk residue (fourth column).  Legend: each control is associated to a mitigation level among three possible values $\Circle=0$ (the control has not been selected for implementation), $\LEFTcircle=0.5$ (the control has been selected for implementation but it is only partially effective to mitigate $T$), or $\CIRCLE=1$ (the control has been selected for implementation and it is fully effective to mitigate $T$).}
    \scriptsize
    \begin{tabular}{c|l|c|c}
\begin{tabular}[c]{@{}c@{}}Threats\\ $(\mathcal{T})$\end{tabular} &
\multicolumn{1}{c|}{\begin{tabular}[c]{@{}c@{}}Controls\\$\{\mathcal{C}_T\}_{T \in \{T_1,T_2,T_3,T_4,T_5\}}$\end{tabular}}&
  \begin{tabular}[c]{@{}c@{}}Mitigation\\ Mapping\\ $\mu_T$\end{tabular} &
  \begin{tabular}[c]{@{}c@{}}Risk\\ residue\\ $x_T$\end{tabular} \\\hline\hline
\multirow{5}{*}{$T_1$}  
& $c_1$) Purpose specification  &\CIRCLE  & \multirow{5}{*}{0.4}\\
& $c_2$) Ensuring limited data processing        &\CIRCLE   &                  \\
& $c_3$) Ensuring purpose related processing     &\LEFTcircle &                  \\
& $c_4$) Ensuring data minimization              &\LEFTcircle &                  \\
& $c_5$) Enabling data deletion                  &\Circle &                 \\\hline 
\multirow{10}{*}{$T_2$} 
& $c_6$) Ensuring data subject authentication    &\CIRCLE& \multirow{10}{*}{0.35} \\
& $c_7$) Ensuring staff authentication           &\CIRCLE&                  \\
& $c_8$) Ensuring device authentication          &\LEFTcircle&                  \\
& $c_9$) Logging access to personal data         &\LEFTcircle&                  \\
& $c_{10}$) Performing regular privacy audits       &\Circle &                  \\
& $c_{11}$) Ensuring data anonymization             &\LEFTcircle&                  \\
& $c_{12}$) Providing confidential communication    &\CIRCLE&                  \\
& $c_{13}$) Providing usable access control         &\LEFTcircle&                  \\
& $c_{14}$) Ensuring secure storage                 &\CIRCLE&                  \\
& $c_{15}$) Ensuring physical security             &\LEFTcircle&                  \\\hline 
\multirow{4}{*}{$T_3$}  
& $c_{16}$) Providing confidential communication    &\CIRCLE& \multirow{4}{*}{0.25}\\
& $c_{17}$) Logging access to personal data         &\LEFTcircle&                  \\
& $c_{18}$) Ensuring data subject authentication    &\CIRCLE&                  \\
& $c_{19}$) Ensuring data anonymization             &\LEFTcircle&                  \\\hline 
\multirow{3}{*}{$T_4$}  
& $c_{20}$) Enabling offline authentication         &\Circle  & \multirow{3}{*}{0.83}\\
& $c_{21}$) Network monitoring                      &\LEFTcircle &                  \\
& \begin{tabular}[c]{@{}l@{}}
$c_{22}$) Prevention mechanisms for DoS attacks like firewalls, etc.\end{tabular}
 &\Circle   &  \\\hline 
\multirow{3}{*}{$T_5$} 
& $c_{23}$) Informing data subjects about data processing   &\LEFTcircle& \multirow{3}{*}{0.66}  \\
& $c_{24}$) Handling data subject’s change requests &\LEFTcircle&                  \\
& $c_{25}$) Providing data export functionality     &\Circle&                 
\end{tabular}
\label{tab:threat-control}
\end{table}
according to~(\ref{eq:msrmp2}) and~(\ref{eq:impact}), the risk residue $x_T\in\{1-m(T)\}$ with $m(T)=1-\frac{\Sigma_{c\in \mathcal{C}} \mu_T(c)}{|\mathcal{C}_T|}$, i.e.\ $x_T$ is the risk residue obtained by applying a certain combination of the security controls available in $\mathcal{C}_T$ for the threat $T$ according to the mitigation mapping $\mu_T$. Recall also that $\mu_T(c)$ measures the impact of $T$ after applying control $c\in \mathcal{C}_T$ and thus $m(T)$ measures the aggregated mitigating effect of selecting a given set of controls in $\mathcal{C}_T$ on the risk of $T$ materializing (under the assumption that the mitigations are independent of each other).  The third and fourth columns of Table~\ref{tab:threat-control} show a given mitigation mapping $\mu_T$ and the associated value $x_T$ of the resulting risk residue.  It will be the task of an automated solver to explore the space of all possible values of $x_T$ and find those that are Pareto-optimal solutions of the MSRMP instance (\ref{eq:msrmp2}) so that it is possible to derive the optimal mitigation mappings as described at the end of Section~\ref{subsec:problem}; see Section~\ref{subsec:il-1st-attempt}).  As already said above, we will see that finding optimal values for $x_T$ is crucial also for solving instances of the general problem statement (\ref{eq:msrmp}); see Section~\ref{subsec:il-refinement}.

\subsection{\textbf{Defining Impacts Levels According to Stakeholders: A First Attempt}}
\label{subsec:il-1st-attempt}
Different stakeholders have different criteria that define what they consider risky. Data controllers (e.g., companies) typically choose business impact criteria, such as financial impact or reputation, whereas data subjects (e.g., individuals) evaluate risk based on impact on their personal sphere. % Therefore, each stakeholder has different protection criteria to be concerned about the impact of a threat.
For the running example introduced in Section~\ref{subsec:scenario}, we consider the \textit{social situation, individual freedom, financial situation}~\cite{ref_article4}, and \textit{health condition} as the data subject protection criteria while for the data controller, \textit{reputational situation} and \textit{financial situation} are the protection criteria, which are linked to indirect or direct pecuniary losses.  
Additionally, each stakeholder has different preferences, which result in different importance given to different criteria; e.g., in the running example, the health condition criterion is more momentous than others for patients. We capture these high-level stakeholder preferences by assigning a weight to each stakeholder’s protection criterion.  The associations among stakeholders, protection criteria, and weights are shown in the first three columns of Table~\ref{tab:threat-preference-association}.  Formally, we assume the availability of a set $\mathcal{P}$ of protection criteria, a family $\{\mathit{PW}_p^s\}_{p\in \mathcal{P}, s\in \mathcal{S}}$ of weights associated to a preference $p$ for each stakeholder $s$ besides the definitions of $\mathcal{T}$, $\{\mathcal{C}\}_{T\in \mathcal{T}}$, and $x_T$ for $T\in \mathcal{T}$ as discussed above in this section.  

The additional information in $\mathcal{P}$ and $\{\mathit{PW}_p^s\}_{p\in \mathcal{P}, s\in \mathcal{S}}$ are used to define the impact level $i_s$ by giving a quantitative evaluation of the negative influence that a threat $T\in \mathcal{T}$ may have on a preference $p\in \mathcal{P}$ for a certain stakeholder $s\in \mathcal{S}$.  The intuition is to characterize how each threat is perceived as more or less dangerous by each stakeholder in relation to his/her own protection criteria.  For instance, in the context of the running example, it is very unlikely that excessive storage of patients’ health data would damage the data controller’s reputation; by increasing stored data, there is financial damage on the data controller cause of cost of storage and management of the IT infrastructure.  On the other hand, the reputation of patients is not affected by excessive storage of personal data; indeed, a larger amount of stored data increases the impact of data breaches and leaks on the rights and freedoms of patients.  For this, we assign an impact value in $\mathcal{IL}$ (recall that this set typically contains a finite set of integer values from $0$ to $4$ included) to the level of aversion that each stakeholder $s$ has for a threat $T$ acting on a given protection criterion $p$. Formally, we assume the definition of an \emph{aversion mapping} $\mathit{al}_p^s:\mathcal{P}\to \mathcal{IL}$ for each preference $p\in \mathcal{P}$ and stakeholder $s\in \mathcal{S}$.  At this point, we are in the position to define $i_s$ by combining the weight $\mathit{PW}_p^s$ and the mapping $\mathit{al}_p^s$ as follows: 
% [[[MAJID, I HAVE ADDED THE TERM $\frac{1}{|\mathcal{T}|}$ TO THE FOLLOWING FORMULA OTHERWISE, THERE SEEMS TO BE A MISMATCH WITH THE EXAMPLE THAT FOLLOWS, PLEASE CHECK!!!]]]
\begin{equation} \label{eq:threatImpact} 
   i_s(T)= \frac{1}{|\mathit{il}_{max}|} \sum_{p\in \mathcal{P}} \mathit{al}_p^{s}(T) ~\times ~ PW_{p}^{s} 
\end{equation}
where $\mathit{il}_{max}\in \mathcal{IL}$ represents the maximum impact level (in our case, it is 4). 
The crux to specify $i_s$ is thus to define the family $\{\mathit{al}_p^s\}_{p\in \mathcal{P}, s\in \mathcal{S}}$ of aversion mappings.  This can be done as shown in the fourth column of Table~\ref{tab:threat-preference-association} where each threat $T\in \mathcal{T}$ gets an aversion level $\mathit{al}_p^s$ between $0$ and $4$ (recall that 0 means no, 1 low, 2 moderate, 3 critical, and 4 catastrophic impact) for each protection criterion $p$ and stakeholder $s$. 
%This value intends to capture the contribution of that specific protection criterion to the overall evaluation of the risk impact. 
Intuitively, the values are assigned by answering the question “For the stakeholder $s$, what would be the impact level on the criterion $p$ if the threat $T$ happen?”
%To evaluate the impact level of the threats, each protection criterion of stakeholders will be assessed under this question “What would be the impact level on the ... if the threat happen?”.
%In Table~\ref{tab:threat-preference-association}, we have assigned an impact value in a scale from 0 to 4 (0-no impact, 1-low, 2-moderate, 3-critical, 4-catastrophic) to each stakeholder’s protection criterion for each $T \in \mathcal{T}$. We have assigned a weight to each protection criterion (PW), as can be seen in the third column of Table~\ref{tab:threat-preference-association}.
To illustrate, consider Table~\ref{tab:threat-preference-association} in which the aversion level of the \textit{health condition} for the second threat ($T_2$) according to the data subject ($s=\mathit{DS}$) is 4 and thus the value of $i_{\mathit{DS}}(T_2)$ will be $\frac{(0.4 \times 4)+(0.2 \times 2)+(0.3 \times 2)+(0.1 \times 3)}{4}=0.725$ according to (\ref{eq:threatImpact}).

% This step’s output will be the list of stakeholders and their preferences and an assigned weight to each of them, which is the input for the next process.

\begin{table}
\caption{The assigned impacts to each stakeholders' preferences for each threat in our scenario.}
\label{tab:threat-preference-association}
\resizebox{\textwidth}{!}{\begin{tabular}{c|l|c|ccccc}

\multirow{2}{*}{Stakeholders ($\mathcal{S}$)} &
  \multicolumn{1}{c|}{\multirow{2}{*}{Protection Criteria ($\mathcal{P}$)}} &
  \multirow{2}{*}{\begin{tabular}[c]{@{}c@{}}Weights\\ $\mathit{PW}_p^s$\end{tabular}} &
  \multicolumn{5}{c}{Aversion level ($\mathit{al}_p^s$)} \\ \cline{4-8}  & \multicolumn{1}{c|}{}  &     &$T_1$ &$T_2$ & $T_3$&$T_4$&$T_5$ \\ \hline \hline
\multirow{4}{*}{Data Subject}    & Health   condition     & 0.4 & 0  & 4  & 0  & 3  & 4  \\ 
                                 & Individual freedom     & 0.2 & 0  & 2  & 4  & 3  & 3  \\ 
                                 & Social situation       & 0.3 & 1  & 2  & 3  & 0  & 3  \\ 
                                 & Financial   situation  & 0.1 & 0  & 3  & 1  & 0  & 3  \\ \hline
\multirow{2}{*}{Data Controller} & Reputational situation & 0.4 & 1  & 2  & 3  & 2  & 2  \\  
                                 & Financial   situation  & 0.6 & 2  & 2  & 3  & 3  & 2  \\ 
\end{tabular}}
\end{table}

% impact evaluation process
%By having the obtained \textit{IV} values from the association between threats and protection criteria, and the \textit{weight} assigned to each criterion (i.e., PW), the impact level of each threat for each stakeholder can be computed through following formula:  
% The computed impact levels will be used as one of the inputs for the risk calculation process.
% Formula~~\ref{eq:threatImpact} computes the impact level of each threat for each stakeholder.
%\begin{equation} \label{eq:threatImpact} 
%  i_s(T)= \sum_{\substack {p\in \mathcal{P}\\ }} IV_{(T,p)}^{s} ~\times ~ PW_{p}^{s}
%\end{equation}
%In this formula, $IV_{(T,p)}^{s}$ represents the impact value of threat $T$ for preference $p$ from the point of view of stakeholder $s$, and $PW_{P}^{s}$ is the assigned weight to that preference. 

To summarize, we have described an approach to define $i_s$ by assuming the capability of identifying protection criteria for each stakeholder (i.e.\ being able to define the set $\mathcal{P}$), of quantifying the relevance of each such criterion (in a scale between $0$ and $1$) for each stakeholder (i.e.\ being able to define the family $\{\mathit{PW}_p^s\}_{p\in \mathcal{P}, s\in \mathcal{S}}$), and assigning an aversion level of each stakeholder when a  threat impacts a given protection criterion (i.e.\ defining the family $\{\mathit{al}_p^s\}_{p\in \mathcal{P}, s\in \mathcal{S}}$).  This allows us to define an instance of the MSRMP (\ref{eq:msrmp2}) which, as we will see in the following, can be solved by using available techniques and then, as described at the end of Section~\ref{subsec:problem}, to identify the set of Pareto optimal mitigation mappings that minimize the risks with respect the various stakeholders.  However, we observe that it may be non-obvious to quantify the weights in $\{\mathit{PW}_p^s\}_{p\in \mathcal{P}, s\in \mathcal{S}}$ and the aversion level mappings in $\{\mathit{al}_p^s\}_{p\in \mathcal{P}, s\in \mathcal{S}}$ as their definitions are quite subjective for each stakeholder.  This is somehow unavoidable because it is up to each stakeholder to define $i_s$, however it is important to mitigate possible bias that would make the solutions of the corresponding instance of the MSRMP (\ref{eq:msrmp2}) hardly useful in practice or even detrimental because of an over or under estimation of the risk levels with negative business or privacy impacts, respectively, on some stakeholders.  We can consider to assign the definitions of  $\{\mathit{PW}_p^s\}_{p\in \mathcal{P}, s\in \mathcal{S}}$ and $\{\mathit{al}_p^s\}_{p\in \mathcal{P}, s\in \mathcal{S}}$ to two independent groups of experts for each stakeholder so to mitigate possible bias.  In the next section, we describe a refined approach to define an instance of the MSRMP (\ref{eq:msrmp}) that aims to further reduce the level of subjectivity of each stakeholder in defining $i_s$. 
%For example, the data subject's \emph{health condition} has the highest weight among all of the data subject's preferences whereas  the greatest weight from the data controller point of view is \emph{financial situation}. It is worth restating that the reason for using such a cross-weighting method is to decrease the level of subjectivity in the assessment.
%By referring to Section~\ref{subsec:problem}, by having the two values $x_T$ and $i_s(T)$, we were able to compute the overall impact residue (i.e., $oir(s)$). To do this, here, we use the same formula to obtain $x_T$ and defined Formula \ref{eq:threatImpact} to compute $oir(s)$. However, notice that this heavily depends on each stakeholder, which results in being too subjective in risk impact evaluation. 

\subsection{\textbf{A Less Subjective Definition of Impact Levels}}
\label{subsec:il-refinement}

Our goal is to reduce the level of subjectivity with which $i_s$ is defined.  The idea is to refine the definition of $i_s$ given above by introducing a cross-weighting system to reduce bias resulting from stakeholders as much as possible.  Besides the availability of a set $\mathcal{P}$ of protection criteria and a family $\{\mathit{PW}_p^s\}_{p\in \mathcal{P}, s\in \mathcal{S}}$ of weights associated to a preference $p$ for each stakeholder $s$, we consider a set $\mathcal{G}$ of protection goals which play a crucial role in identifying appropriate security controls (see, e.g.,~\cite{zwingelberg2011}).  Indeed, Confidentiality, Integrity, and Availability are obvious candidates to be included in the set $\mathcal{G}$ (see, e.g.,~\cite{brooks2017}).  However, these are not enough to consider the complex protection requirements deriving from national and international legal provisions such as those concerning data protection contained in the GDPR.  For this reason, in the rest of the paper, we assume the set $\mathcal{G}$ to contain the ``data protection goals" introduced by the Standard Data protection Model (SDM)~\cite{fur2017standard}.

To systematize data protection requirements of the GDPR, the SDM employs ``protection goals". The data protection requirements seek to ensure legal compliance processing, which technological and organizational safeguards must ensure. The assurance consists in lowering the risk of deviations from legally compliant processes to a suitable degree. Unauthorized processing by third parties and the failure to carry out mandatory processing procedures are examples of deviations to avoid. The data protection goals combine and arrange the criteria for data protection requirements and can be operationalized through integrated, scalable measures~\cite{fur2017standard}. These protection goals are 
\begin{enumerate}[{G}1.]
\item \textit{\textbf{Confidentiality}} refers to the requirement that no person is allowed to access personal data without authorisation.
\item \textit{\textbf{Integrity}} refers, on the one hand, to the requirement that information technology processes and systems continuously comply with the specifications that have been determined for the execution of their intended functions. On the other hand, integrity means that the data to be processed remain intact, complete, and up-to-date.
\item \textit{\textbf{Availability}} is the requirement that personal data must be available and can be used properly in the intended process. Thus, the data must be accessible to authorised parties and the methods intended for their processing must be applied.
\item \textit{\textbf{Unlinkability $\&$ Data minimization}} where the \emph{unlinkability} goal refers to the requirement that data shall be processed and analysed only for the purpose for which they were collected, while the \emph{data minimization} goal covers the fundamental requirement under data protection law to limit the processing of personal data to what is appropriate, substantial and necessary for the purpose. 
\item \textit{\textbf{Transparency}} refers to the requirement that the data subject as well as the system operators and the competent supervisory authorities can identify to a varying extent, which data are collected and processed for a particular purpose, and which systems and processes are used for this purpose, where the data flow to which purpose, and who is legally responsible for the data and systems in the various phases of data processing.
\item \textit{\textbf{Intervenability}} refers to the requirement that the data subjects are effectively granted the right to notification, information, rectification, blocking and erasure at any time.
\end{enumerate}
%----refined
The SDM have provided precise mappings between the GDPR requirements and these protection goals (for more details, see the table~\footnote{The Standard Data Protection Model (SDM), \url{https://www.datenschutzzentrum.de/uploads/sdm/SDM-Methodology_V2.0b.pdf}} on pages 28 to 30). These mappings can be interpreted as if threats adversely affecting these protection goals mean non-compliance with the GDPR requirements.
Working with protection goals simplifies the modeling of functional requirements in use cases and the visualization of conflicts. They also enable the methodical application of legal requirements into technological and organizational measures and are therefore ``optimization requirements".
We observe that our approach can be applied with other protection goals, we consider those of~\cite{fur2017standard} only for the sake of concreteness.

The goal of the approach discussed below is twofold: (i) identify how many goals each threat is impacting and (ii) measure the amplitude of the impact on each goal of a given threat.  We start by considering (i).  
%
%In risk management, the impact assessment is a critical task because it aims to estimate the potential consequences of threats and how stakeholders evaluate them. 
%It is mostly a subjective evaluation, which makes it particularly critical when attempting to estimate the impact from third parties' point of view, such as the data subject.
%We accept this subjectivity level as unavoidable, but we mitigate it by conceiving a cross-weighting system, making impact assessment more systematic.
% table 3+6 formula 9-10-11
% Threat-Goal
\begin{table}\center
\caption{Affected protection goals by each threat and the observation weights in our scenario, G1= Confidentiality, G2= Integrity, G3= Availability, G4= Unlinkability $\&$ Data minimization, G5= Transparency, and G6= Intervenability. }
\footnotesize
\label{tab:Threat-Goals}
\resizebox{\textwidth}{!}{\begin{tabular}{c|cccccc|c}
\multicolumn{1}{c|}{\multirow{2}{*}{Threat}} &
  \multicolumn{6}{c|}{Data Protection Goals} &
  \multirow{2}{*}{\begin{tabular}[c]{@{}c@{}}Observation\\ Weights (OW)\end{tabular}} \\ \cline{2-7}
\multicolumn{1}{c|}{} & G1 & G2 & G3 & G4 & G5 & G6 &      \\ \hline \hline
$T_1$ & \ding{53}  & -  & -  & \ding{53}  & -  & -  & 2/10 \\ 
$T_2$ & \ding{53}  & \ding{53}  & \ding{53}  & -  & -  & -  & 3/10 \\ 
$T_3$ & \ding{53}  & -  & -  & \ding{53}  & -  & -  & 2/10 \\ 
$T_4$ & -  & \ding{53}  & \ding{53}  & -  & -  & -  & 2/10 \\ 
$T_5$ & -  & -  & -  & -  & -  & \ding{53}  & 1/10 \\ 
\end{tabular}}
\end{table}
%As we observed earlier in Section~\ref{sec:Back}, six protection goals have been specified by the SDM~\cite{fur2017standard}. 
%These protection goals reflect the protection objectives in information security and present the data subject’s perspective, whose rights are at stake when considering the GDPR.  
%Generally, each threat has potentially an impact on one or more protection goals, which depends on the threat’s nature. 
For example, a “\emph{Denial of service}” threat will intuitively have more impact on the data availability goal rather than on the integrity goal; an “\emph{Identity theft}” threat will have more impact on the data confidentiality goal.  To keep track of this, we use a \emph{Threat-Protection Goals association} as shown in the first two columns in Table~\ref{tab:Threat-Goals}   %  as action of mapping the data protection goals to the list of threats in $\mathcal{T}$.
% as derived by the threat scenario identification process in Figure~\ref{fig:methodology}.       
%The association between these protection goals and threat in $\mathcal{T}$
%is shown in Table~\ref{tab:Threat-Goals}, 
where the “\ding{53}” (“-”) mark in a cell means the goal in the column is affected (not affected, respectively) by the threat in the row (the particular instance of the threat-protection goals association is related to the running example of Section~\ref{subsec:scenario}).  Intuitively, the more a threat impacts multiple goals, the more it is considered pervasive (e.g., threat $T_2$ is the most pervasive in Table~\ref{tab:Threat-Goals} as it affects 3 goals); the more a goal is impacted by multiple threats, the more it is considered scattered (e.g., goal $G1$ is the most scattered in Table~\ref{tab:Threat-Goals} as it impacts 3 threats).  The third column of Table~\ref{tab:Threat-Goals} shows the so called Observation Weight 
\begin{equation} \label{eq:OW}  
   OW_{T}= \frac{AG_{T}}{{\sum_{T \in \mathcal{T}}AG_T}} 
\end{equation}
that measures how much a threat $T$ is pervasive for the goals in $\mathcal{G}$, where $\mathit{AG}_T$ is the number of goals in $\mathcal{G}$ affected by a threat $T\in \mathcal{T}$.  For example, in Table~\ref{tab:Threat-Goals}, the observation weight $OW_{T_1}$ is $2/10$, where $G1$ and $G4$ are the two affected goals by $T_1$, and the total number of affected goals is 10. 

We now consider objective (ii), namely to measure the amplitude of the impact on each goal of a given threat.  This is necessary as soon as we realize that the information in Table~\ref{tab:Threat-Goals} is not enough alone to define $i_s$ because it may be the case that the impact value can be much higher when a goal is impacted severely by a single threat rather than when this is impacted by many threats but only lightly.  
%This information is used in combination with others to derive the less subjective overall impact. We derive the “\textit{Observation Weight}” for each threat according to the number of protection goals affected by the threat with respect to all affected protection goals in total. For simplicity, we assume all the protection goals have equal importance. The observation weights compute through the following formula:
%\begin{equation} \label{eq:OW}  
%   OW_{T}= \frac{NG_{T}}{{\sum_{T \in \mathcal{T}}NG_T}} ~~~ for ~~T \in \mathcal{T}
%\end{equation}
%where $NG_{T}$ represents the number of affected goals by threat $T \in \mathcal{T}$.
%
We do this in two steps.  First, we define the \emph{normalized threat criticality level} as 
\begin{equation} \label{eq:Sc}  
    NTC_T= \frac{OW_T \times x_T}{\sum_{T \in \mathcal{T}} (OW_T \times x_T)}
\end{equation}
to quantify the severity of a threat $T\in \mathcal{T}$ (recall that $x_T$ is the impact residue of the threat $T$ after applying the security controls according to a mitigation mapping $\mu_T$).
Intuitively, $\mathit{NTC}_T$ is the level of danger of a threat $T$ among all threats in $\mathcal{T}$, or in other words, the relative importance of $T$ with respect to all other threats in $\mathcal{T}$. 
%how much a threat has precedence over the other threats
% We define one further value with purpose of quantifying the severity of threats, and it is called \emph{threat criticality (TC)}.
% The purpose of threat criticality ($TC$) is to quantify the seriousness of each threat.
%We use a simple technique to evaluate $TC$ based on the relationship between the protection goals and the mapped controls for each threat (i.e., $\{\mathcal{C}_T\}_{T \in \mathcal{T}}$). 
%To quantify the threat criticality value for each threat, two values must be considered, (i) the observation weight ($OW$), and (ii) the risk residue ($x_T$). Basically, the threat criticality shows the level of danger of a threat among all threats, or in other words, how much a threat has precedence over the other threats. Through the following formulas, the threat criticality values compute and then normalize:
% The threat criticality computes by Formula~\ref{eq:Sc} to determine how much a threat has precedence over the other threats. 
%\begin{equation} \label{eq:Sc}  
%    TC_T= OW_T \times x_T 
%\end{equation}
%\begin{equation} \label{eq:RSP}
%  \forall_{T=1,..,t} \, \: 
% NTC_T=\frac{TC_T}{{\sum_{T \in \mathcal{T}}TC_T}}
%\end{equation}
\begin{table}[!b] \center
        \caption{Threat criticality and impact level values together with the computed protection goals' impacts for each threat for the data subject (DS) and the data controller (DC).} \label{tab:risk-goals}
	    \includegraphics[width=0.99\textwidth]{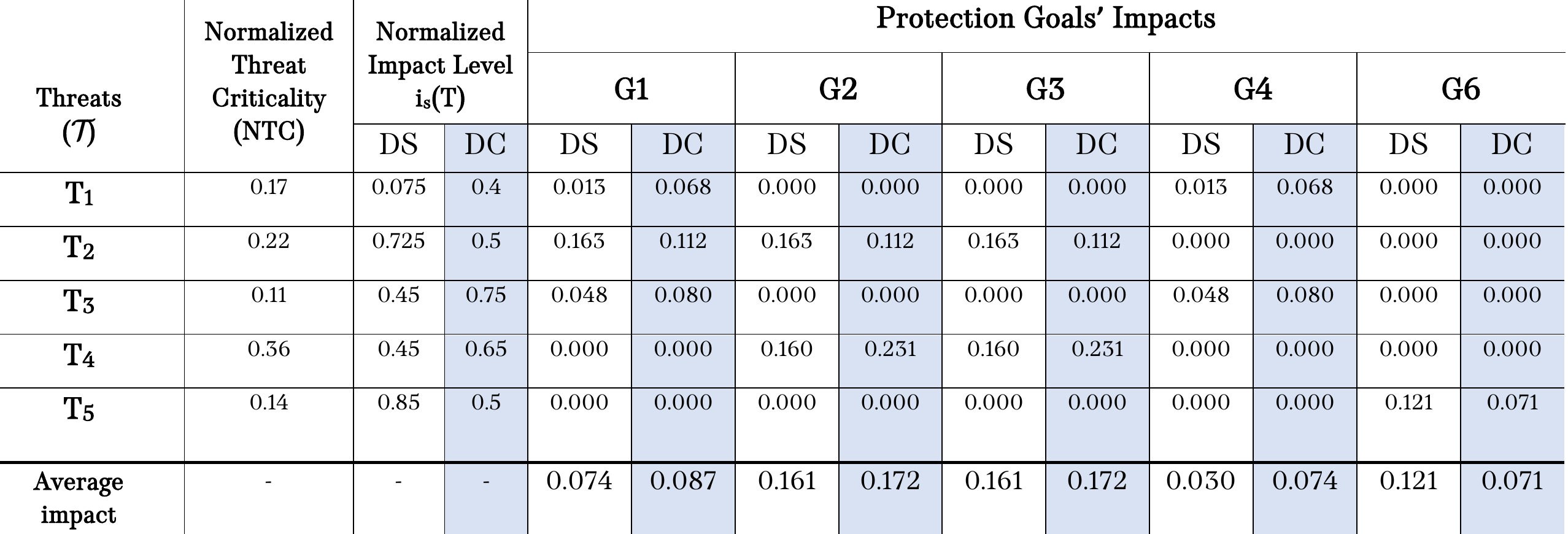}
\end{table} 
% We then normalize the obtained values through Formula~\ref{eq:RSP} to calculate risk exposures. 
% The normalized threat criticality values for $T \in \mathcal{T}$ are shown in the second column of Table~\ref{tab:risk-goals} by using the obtained observation weights (Table~\ref{tab:Threat-Goals}) and the calculated $TE$ values (Table~\ref{tab:threat-control}). 
By having  obtained the observation weights (in Table~\ref{tab:Threat-Goals}) and the calculated $x_T$ values (in Table~\ref{tab:threat-control}), the computed normalized threat criticality values for $T \in \mathcal{T}$ are shown in the second column of Table~\ref{tab:risk-goals}. 

The second step to achieve goal (ii) above is to use the normalized threat criticality level to weight the function $i_s$ defined in Section~\ref{subsec:il-1st-attempt} when considering a certain protection goal $G\in \mathcal{G}$ for a given stakeholder $s$ so to define the overall impact residue as follows
\begin{equation} \label{eq:risk-exposure}
    	\mathit{oir}(s) =  \sum_{G\in \mathcal{G}} \left(\frac{ ~\sum_{T\in\mathcal{T}} \xi_{T,G} \times \mathit{NTC}_{T} \times i_s(T)}{\#(\mathcal{T},{G})}\right)
\end{equation} 
where $\xi_{T,G}$ is $1$ when the threat $T\in \mathcal{T}$ compromises the goal $G\in \mathcal{G}$ and $0$ otherwise; $\#(\mathcal{T},{G})$ is the number of threats in $\mathcal{T}$ that have an impact on the goal $G$ (this means that $\#(\mathcal{T},{G})=\sum_{T\in \mathcal{T}} \xi_{T,G}$). Observe that the expression between parentheses in (\ref{eq:risk-exposure}) can be seen as the average impact on a given goal $G$ with respect to the threats in $\mathcal{T}$ that are relevant to $G$.  
For instance, according to Table~\ref{tab:Threat-Goals}, the \textit{intervenability} goal ($G6$) is affected only by $T_5$ which means that $\#(\mathcal{T},{G6})$ is 1.
% As an instance of the risk calculation process, we have reported the risk exposure results in Table~\ref{tab:risk-goals} for the case of when $TE$ values are the ones reported in Table~\ref{tab:threat-control}. 
According to Table~\ref{tab:risk-goals}, the average impact  of the \emph{confidentiality} goal ($G1$) for the data subject is $0.074$, while the same value for the data controller is $0.087$. 
Finally, observe that since the \emph{transparency} goal ($G5$) is not affected by anyone of the threats (according to Table~\ref{tab:Threat-Goals}), it is not mentioned in Table~\ref{tab:risk-goals} neither used for calculating the overall impact residue.
By aggregating the impact average of protection goals, the overall impact residue from the data subject’s point of view is $\mathit{oir}(\mathit{DS})=0.549$, and for the data controller is $\mathit{oir}(\mathit{DC})=0.576$. 

At this point, we  are in the position to define instances of the MSRMP statement (\ref{eq:msrmp}) by using (\ref{eq:risk-exposure}) as the definition of the overall impact residue rather than those proposed in Section~\ref{subsec:problem}.    We also observe that by substituting the definition (\ref{eq:Sc}) to $\mathit{NTC}_{T}$ in the expression of $\mathit{oir}(s)$, it is easy to see that we can derive a MSRMP similar to (\ref{eq:msrmp2}), i.e.\ considering $x_T$ as variables rather than $\mu_T$ for $T\in \mathcal{T}$, for which it is possible to apply the same technique discussed at the end of Section~\ref{subsec:problem} that allows us to solve an optimization problem over a smaller search space and then derive optimal solutions for the original problem.

% \section{Discussion}

%We also normalized $i_s(T)$ values by bringing values in the range of [0,1] as they are reported in the third column of Table~\ref{tab:risk-goals}.
% The normalized threat criticality values for each threat in our scenario by using the obtained observation weights (Table~\ref{tab:Threat-Goals}) and the calculated $TE$ values (Table~\ref{tab:threat-control}) are reported in the second column of Table~\ref{tab:risk-goals}.
% Risk Calculation
%To calculate overall impact residue (i.e., oir(s)), firstly, we compute the goal impact average (GIA), which represents the fact of how much the protection goals are at risk. The goal impact average for $G \in \mathcal{G}$ computes through Formula~\ref{eq:risk-exposure} for each stakeholder, and then by aggregating them, the overall impact residue will obtain.

%\begin{equation} \label{eq:risk-exposure}
%    	GIA_{s}^{G}= \frac{ ~\sum_{T\in\mathcal{T}}(if~T~compromises~G~then) ~NTC_{T} \times i_s(T)}{X_{G}}
%\end{equation} 
% After obtaining the two values $NTC$ and normalized $i_s(T)$, we calculate the risk levels through Formula~\ref{eq:risk-exposure} for each stakeholder in terms of how much the protection goals are at risk. Then by aggregating them, the final risk levels or, in other words, the obtained objective functions from a multi-objective point of view.
%where $X_{G}$ represents the total number of affection of the goal $G$ by threats, in another word, how many threats affect the goal $G$. 

\section{Implementation and Experimental Evaluation}
\label{sec:impl-exp}

To validate the applicability of the proposed methodology, we have implemented a tool % (in Java with JSON as the data representation format) 
able to assist in defining an instance of the MSRMP as discussed in Section~\ref{sec:new-methodology} and performed two sets of tests in order to experimentally evaluate the practicality of our approach~\footnote{The code of the tool and the material to replicate the experiments are available at~\url{https://github.com/stfbk/MSRMP}}.

\begin{figure}[!t]
    \centering
	\includegraphics[width=0.95\textwidth]{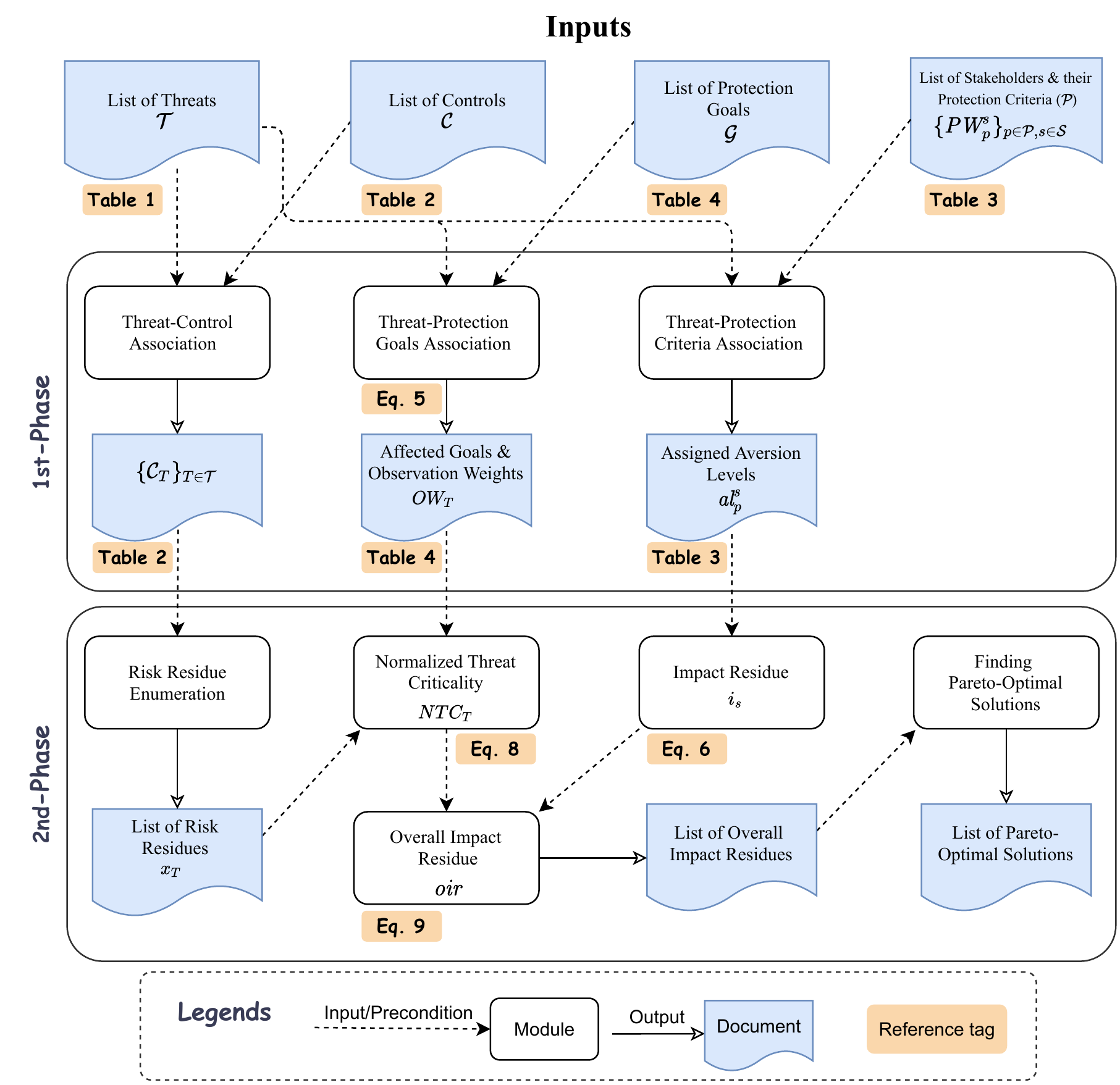}
    \caption{Architecture of the implemented tool}
    \label{fig:methodology}
\end{figure}

The goal of the tool is two-fold, namely (i) assisting in the definition of an instance of the MSRMP and (ii) automatically solving the resulting instance.  
The architecture of the tool is illustrated in Figure~\ref{fig:methodology}; the modules are implemented in Java while the documents use JSON as the data representation format.  The tool operates in two phases (see outer boxes in the figure) and assumes the availability of the sets of stakeholders $\mathcal{S}$, threats $\mathcal{T}$, security controls $\mathcal{C}$, protection criteria $\mathcal{P}$ together with their weights $\{\mathit{PW}_p^s\}_{p\in \mathcal{P}, s\in \mathcal{S}}$, and goals $\mathcal{G}$; the first three are discussed in Section~\ref{subsec:problem}, the fourth in Section~\ref{subsec:il-1st-attempt}, and the last in Section~\ref{subsec:il-refinement}.  The architecture also reports how tabular definitions of the various entities can be given; for instance, the set $\mathcal{T}$ of threats can be defined as in Table~\ref{tab:threat-table} and the set $\mathcal{P}$ of protection goals together with their weights $\{\mathit{PW}_p^s\}_{p\in \mathcal{P}, s\in \mathcal{S}}$ as in Table~\ref{tab:threat-preference-association}.  We assume that these inputs are derived from the application of available and well-known techniques for risk assessment as already discussed above; our approach is agnostic with respect to the particular methodology used.  The tables specifying the inputs above are encoded in JSON format.   

The first phase is semi-automated and a preliminary step to the definition of an instance of the MSRMP.  More precisely, it defines the association between controls and threats $\{\mathcal{C}_T\}_{T\in \mathcal{T}}$ (see Section~\ref{subsec:problem}), the aversion level mapping $\mathit{al}_p^s$ for each protection criteria $p\in \mathcal{P}$ and stakeholder $s\in \mathcal{S}$ (see Table~\ref{tab:threat-preference-association} in Section~\ref{subsec:il-1st-attempt}), and the observation weight $\mathit{OW}_T$ for each threat $T\in \mathcal{T}$; see last column of Table~\ref{tab:Threat-Goals} whose value is derived according to (\ref{eq:OW}).  The first two outputs of this phase are obtained with human intervention as the user needs to identify which security controls are effective for each threat and which is the level of aversion of each stakeholder for a given protection criteria to be violated whereas the last one is automatically derived after the user has specified which goals are affected by each threat.  % All the outputs of the phase are represented in JSON format.  

The second phase is fully automated and aims to define and solve an instance of the MSRMP.  This requires to use the outputs of the first phase to define the impact level mapping $i_s$ for each stakeholder $s\in \mathcal{S}$; see Section~\ref{subsec:problem}) along the lines of Section~\ref{subsec:il-1st-attempt} and then the overall impact residue $\mathit{oir}$ as discussed in Section~\ref{subsec:il-refinement}.  At this point, the tool has fully defined an instance of the MSRMP (\ref{eq:msrmp}) and it is left with the task of solving it.  For this, it needs to enumerate all risk residues $x_T$ for each threat $T\in \mathcal{T}$ by using the approach in Section~\ref{sec:new-methodology} to define Table~\ref{tab:threat-control}, derive the Normalized Threat Criticality values for the various threats, and then adapt the strategy discussed at the end of Section~\ref{subsec:problem} to identify the mitigation mappings that are Pareto optimal. 
% (as above, the outputs of the various processing steps, including the set of Pareto Optimal solutions, in this phase are encoded in JSON format).  

We observe that there are multiple possible strategies to combine the enumeration of risk residues and the identification of Pareto optimal values.  For instance, one can first compute the entire set of feasible solutions and only after look for Pareto optimal ones or one can imagine to interleave the two activities by computing the Pareto optimal values in  different subsets of the whole set of feasible solutions and then select those solutions that are Pareto optimal for the entire search space.  Below, we first discuss the computational behavior of the second phase on the running example in Section~\ref{subsec:scenario} and then design two sets of tests to understand which is the most promising strategy to identify the set of Pareto Optimal risk residues or, equivalently, mitigation mappings.  

%--------------Running example Evaluation Section-------

\subsection{\textbf{Applying the Prototype Tool on the Running Example}}
\label{subsec:appl-run-ex}
We discuss the results of applying the second phase of our methodology, as implemented in the prototype tool, on the running example of Section~\ref{subsec:scenario}.  First, the tool computes the whole set of possible solutions whose cardinality is $57,600$; this is as expected from the formula $\Pi_{T\in \mathcal{T}} |{X}_T|= |{X}_{T_1}|\times |{X}_{T_2}|\times |{X}_{T_3}|\times |{X}_{T_4}|\times |{X}_{T_5}|= 10 \times 20 \times 8\times 6 \times 6 = 57,600$ presented in Section~\ref{subsec:problem} (see Example 6). %Therefore, the total set of possible solutions for the running example is $\Pi_{T\in \mathcal{T}} |{X}_T|= |{X}_{T_1}|\times |{X}_{T_2}|\times |{X}_{T_3}|\times |{X}_{T_4}|\times |{X}_{T_5}|= 10 \times 20 \times 8\times 6 \times 6 = 57,600$.

\begin{figure} [!t]
    \centering
	\includegraphics[width=0.95\textwidth]{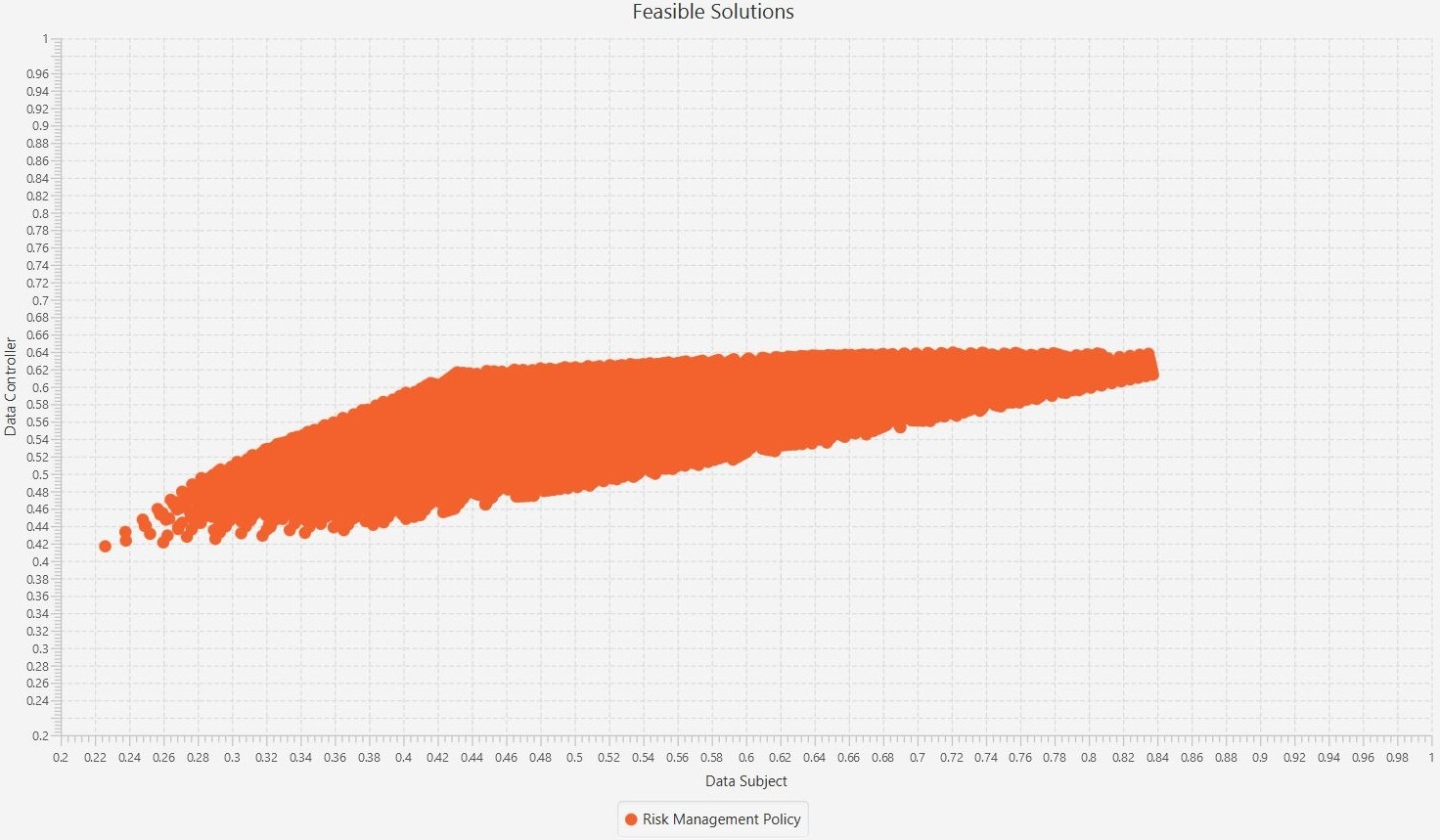}
    \caption{All feasible solutions (i.e., the search space) in the running example scenario.}
    \label{fig:AllPoints}
\end{figure}

% ${X}_{T_1}= \{ \}$
% PLEASE MAJID ADD THE MATHEMATICAL FORMULA TO SHOW THAT THIS IS INDEED THE CORRECT VALUE.  
% 2 seconds computation time, and 435 Mb heap size needed.
% This time is only for generating the all possible solutions. (by using $435$ Mb of RAM heap)
This takes around $2.1$ seconds %$2100$ milliseconds 
on a machine with 16 GB of RAM and a 1.90 GHz CPU. 
Each solution is a pair containing the risk residue values for the Data Subject (DS) and the Data Controller (DC).  Figure~\ref{fig:AllPoints} shows the set of possible solutions plotted on a Cartesian plane whose x-axis shows the risk residue of DS and the y-axis that of DC.  
% Recall that each point is implicitly associated to a set of RMPs whose application gives rise to the corresponding values of the risk residues.  
%In this figure, the data subject’s risk exposure level is plotted on the x-axis, where the y-axis shows the risk exposure level for the data controller.
By looking at the figure, it is immediate to see that the optimal solution is that on the bottom left---whose risk residue values are $0.2260$ for DS and $0.4168$ for DC---% point in the left side bottom where the risk exposure levels for the data subject and the data controller equals 0.2260 and 0.4168, respectively. 
as it dominates all other solutions.  The tool takes around $2.2$ seconds to identify this point as the best one.  

% MAJID I HAVE REWRITTEN THE FOLLOWING TEXT, PLEASE TAKE A LOOK!
After identifying the risk residue levels, one is left with the problem of computing the set of RMPs that generate such values.  A method to do this has been illustrated at the end of Section~\ref{sec:MSRToAP} and implemented in the tool that takes less than $3$ seconds to identify the following tuple
$$\langle x_{T_1}^*, x_{T_2}^*, x_{T_3}^*, x_{T_4}^*, x_{T_5}^* \rangle = \langle 1,0.05,0.125,0.16,0.16 \rangle $$
corresponding to $(0.2260, 0.4168)$ and then to identify all the RMPs associated to the above tuple of $x_T$ values for $T\in \{T_1, T_2, T_3, T_4, T_5\}$.  
%$\mathbb{RS}$ = $\langle x_{T_1}, x_{T_2}, x_{T_3}, x_{T_4}, x_{T_5} \rangle$ =$\langle 1,0.05,0.125,0.16,0.16 \rangle$
By recalling (\ref{eq:mapping-back-eqs}) and that $\mathcal{A}=\{0, 0.5, 1\}$, it is not difficult to see that there are $360=1\times 10\times 4\times 3\times 3$ distinct RMPs associated to the tuple of $x_T$ values above since
\begin{itemize}
    \item there is just one mitigation mapping satisfying 
    $$
    \frac{\Sigma_{c\in \{c_1, ..., c_5\}} \mu_{T_1}(c)}{5} = 1 - x_{T_1}^* = 0
    $$
    as the values in $\mathcal{A}$ are non-negative values;
    \item there are $10$ mitigation mappings satisfying 
    $$
    \frac{\Sigma_{c\in \{c_6, ..., c_{15}\}} \mu_{T_2}(c)}{10} = 1 - x_{T_2}^* = 0.95
    $$
    as the only way to get $9.5$ by adding $10$  values from $\mathcal{A}$ is to have nine of them equal to $1$ and the remaining one to $0.5$; 
    \item there are $4$ mitigation mappings satisfying 
    $$
    \frac{\Sigma_{c\in \{c_{16}, ..., c_{19}\}} \mu_{T_3}(c)}{4} = 1 - x_{T_3}^* = 0.875
    $$
    as the only way to get $3.5=4\times 0.875$ by adding $3$  values from $\mathcal{A}$ is to have three of them equal to $1$ and the remaining one to $0.5$;
    \item there are $3$ mitigation mappings satisfying  
    $$
    \frac{\Sigma_{c\in \{c_{20}, ..., c_{22}\}} \mu_{T_4}(c)}{3} = \frac{\Sigma_{c\in \{c_{23}, ..., c_{25}\}} \mu_{T_5}(c)}{3} = 1 - x_{T_4}^* = 1 - x_{T_5}^* = 0.84 
    $$
    as the only way to get $3.5=3\times 0.8$ by adding three  values from $\mathcal{A}$ is to have two of them equal to $1$ and the remaining one to $0.5$.
\end{itemize}
%At this point, we are left with the problem of configuring the security controls associated to each threat $T\in \{T_1, ..., T_5\}$ so as to obtain the required value $x_T$.  This is a combinatorial problem that can be solved quickly in practice, since the set $\mathcal{C}_T$ of controls mitigating the threat $T$ is usually small. 
%Based on this observations, one of the mitigation mappings associated to the optimal solution is the following:  
%$\mu_{T_2}(c_6)=\mu_{T_3}(c_{16})=\mu_{T_4}(c_{20})=\mu_{T_5}(c_{23})=0.5$, 
%$\mu_{T_2}(c_7)=\mu_{T_2}(c_8)= \mu_{T_2}(c_9)=\mu_{T_2}(c_{10})=\mu_{T_2}(c_{11})= \mu_{T_2}(c_{12})=\mu_{T_2}(c_{13})=
% \mu_{T_2}(c_{14})= \mu_{T_2}(c_{15})=\mu_{T_3}(c_{17})=\mu_{T_3}(c_{18})=\mu_{T_3}(c_{18})=\mu_{T_3}(c_{19})=\mu_{T_4}(c_{21})=\mu_{T_4}(c_{22})=\mu_{T_5}(c_{24})=\mu_{T_5}(c_{25})=1$, and $\mu_T(c)=0$ for each $T\in \{T_1, ..., T_5\}$ and $c\in \{c_1, c_2, c_3, c_4, c_5\}$ (i.e.\ the security controls not occurring as arguments of any mitigation mappings are not implemented). %
The tool mechanizes the observations above and computes the set of security controls associated to the Pareto optimal solutions by solving a variant of the Sum Subset Problem (SSP) in which multisets are considered instead of sets as explained at the end of Section \ref{subsec:problem}.  Indeed, this is so because a mitigation mapping $\mu_T$ associates a control of $\mathcal{C}_T$ with a value in $\mathcal{A} = \{0, 0.5, 1\}$ for each $T\in \mathcal{T}$ and nothing prevents two or more controls to be mapped to the same value in $\mathcal{A}$.  Since all solutions to the SSP should be identified to be able to enumerate all possible mitigation mappings, the algorithm is exponential in the number of security controls associated to each threat, i.e.\ in the cardinality of $\mathcal{C}_T$ for $T\in \mathcal{T}$.  Since such a number is typically low (on average around $5$ and at most $10$ in our experience), the time consumption is quite reasonable in practice being around half a second at most for a single threat $T\in \mathcal{T}$.  To conclude the discussion, the last column of Table~\ref{tab:control-combinations} reports three mitigation mappings associated to the optimal solution considered above. 
Notice that all mitigation mappings associated to the optimal solution above suggest to avoid implementing any security control for threat $T_1$.  This is a consequence of the impact defined in Table \ref{tab:threat-preference-association} that makes $T_1$ relevant only for the social situation of the DS while it is negligible for all other aspects.  Given this remark, one may decide to modify the values to increase the impact of $T_1$ for the DS and then re-run the analysis.  This is a clear advantage of having a high level of mechanization of our methodology.   
%There are other possible mitigation control combinations that will yield $\mathbb{RS}$;
%three such combinations are reported in Table~\ref{tab:control-combinations} (column 3).
%The number of combinations of $\mu_{T}$ values for $T \in \mathcal{T}$ to obtain the identified $x_T$ values above are, $T_1=1, T_2=10, T_3=4, T_4=3, T_5=3$ where the total number of RMP combinations is $360$. 

%MAJID THERE IS NO SET OF RMPs HERE... AT LEAST WE SHOULD INCLUDE A SELECTED SUBSET OF IT FOR ILLUSTRATION AND SAY HOW MANY RMPs ARE IN THE SET...
% PLEASE MAJID ADD THIS NUMBER HERE...

Indeed, the running example is simple and poses no challenges to our prototype implementation.
To understand the scalability of the proposed approach, we have designed a set of synthetic optimization problems whose sets of potential solutions is increasingly large and then experiment with two different strategies to generate and visit such a set in the process of identifying Pareto optimal solutions.  This is reported in Sections~\ref{subsec:experimental-results} below.
\begin{table}[!b] \center
	\caption{Examples of mitigation mappings associated to the optimal solution in Figure \ref{fig:AllPoints}}
    \scriptsize
    \begin{tabular}{c|l|ccc|c}
\begin{tabular}[c]{@{}c@{}}Threats\\ $(\mathcal{T})$\end{tabular} &
\multicolumn{1}{c|}{\begin{tabular}[c]{@{}c@{}}Controls\\$\{\mathcal{C}_T\}_{T \in \{T_1,T_2,T_3,T_4,T_5\}}$\end{tabular}}&
\multicolumn{3}{c|}{\begin{tabular}[c]{@{}c@{}}Possible \\Mitigation \\ Combinations \end{tabular}}&
\begin{tabular}[c]{@{}c@{}}$x_T^*$\\\end{tabular} \\\hline\hline
\multirow{5}{*}{$T_1$} 
&\multirow{1}{*}{$c_1$) Purpose specification}          &\Circle&\Circle& \Circle  & \multirow{5}{*}{1}\\
&\multirow{1}{*}{$c_2$) Ensuring limited data processing}&\Circle&\Circle& \Circle  &  \\
&\multirow{1}{*}{$c_3$) Ensuring purpose related processing}&\Circle & \Circle&\Circle&  \\
&\multirow{1}{*}{$c_4$) Ensuring data minimization}    &\Circle &  \Circle & \Circle &   \\
&\multirow{1}{*}{$c_5$) Enabling data deletion}         &\Circle &\Circle &\Circle &                 \\\hline 
\multirow{10}{*}{$T_2$} 
&\multirow{1}{*}{$c_6$) Ensuring data subject authentication}    &\LEFTcircle&\CIRCLE&\CIRCLE& \multirow{10}{*}{0.05} \\
&\multirow{1}{*}{$c_7$) Ensuring staff authentication}      &\CIRCLE&\LEFTcircle&\CIRCLE&\\
&\multirow{1}{*}{$c_8$) Ensuring device authentication}      &\CIRCLE&\CIRCLE&\LEFTcircle&\\
&\multirow{1}{*}{$c_9$) Logging access to personal data}  &\CIRCLE&\CIRCLE&\CIRCLE&  \\
&\multirow{1}{*}{$c_{10}$) Performing regular privacy audits}   &\CIRCLE  &\CIRCLE &\CIRCLE &    \\
&\multirow{1}{*}{$c_{11}$) Ensuring data anonymization}   &\CIRCLE&\CIRCLE&\CIRCLE&     \\
&\multirow{1}{*}{$c_{12}$) Providing confidential communication} &\CIRCLE&\CIRCLE&\CIRCLE&     \\
&\multirow{1}{*}{$c_{13}$) Providing usable access control}    &\CIRCLE&\CIRCLE&\CIRCLE&    \\
&\multirow{1}{*}{$c_{14}$) Ensuring secure storage}                  &\CIRCLE&\CIRCLE&\CIRCLE&         \\
&\multirow{1}{*}{$c_{15}$) Ensuring physical security}   &\CIRCLE&\CIRCLE&\CIRCLE&       \\\hline 
\multirow{4}{*}{$T_3$}  
&\multirow{1}{*}{$c_{16}$) Providing confidential communication}  &\LEFTcircle &\CIRCLE&\CIRCLE& \multirow{4}{*}{0.125}\\
&\multirow{1}{*}{$c_{17}$) Logging access to personal data}         &\CIRCLE &\LEFTcircle &\CIRCLE&    \\
&\multirow{1}{*}{$c_{18}$) Ensuring data subject authentication} &\CIRCLE&\CIRCLE&\LEFTcircle&                  \\
&\multirow{1}{*}{$c_{19}$) Ensuring data anonymization}   &\CIRCLE &\CIRCLE &\CIRCLE&       \\\hline 
\multirow{3}{*}{$T_4$}  
&\multirow{1}{*}{$c_{20}$) Enabling offline authentication}        &\LEFTcircle &\CIRCLE  &\CIRCLE  & \multirow{3}{*}{0.16}\\
&\multirow{1}{*}{$c_{21}$) Network monitoring}                      &\CIRCLE&\LEFTcircle&\CIRCLE &     \\
&\multirow{1}{*}{$c_{22}$) Prevention mechanisms for DoS attacks like firewalls, etc.}
 &\CIRCLE&\CIRCLE&\LEFTcircle   &  \\\hline 
\multirow{3}{*}{$T_5$} 
&\multirow{1}{*}{$c_{23}$) Informing data subjects about data processing}   &\LEFTcircle&\CIRCLE&\CIRCLE& \multirow{3}{*}{0.16}  \\
&\multirow{1}{*}{$c_{24}$) Handling data subject’s change requests} &\CIRCLE&\LEFTcircle&\CIRCLE&                  \\
&\multirow{1}{*}{$c_{25}$) Providing data export functionality}     &\CIRCLE&\CIRCLE&\LEFTcircle&                 
\end{tabular}
\label{tab:control-combinations}
\end{table}
% -----commented from here 
Preliminarily, we discuss a variant of the MSRMP that, with little effort, can be solved by a minor modification to our approach.  Such a variant is a constrained version of the MSRMP whereby it is possible to identify lower bounds for risk residue levels of the DS and DC, i.e.\ the stakeholders may be willing to accept a risk residue above a certain threshold according to their risk appetite, i.e.\ the amount of risk the stakeholder is willing to take in pursuit of objectives it considers valuable.  In other words, the set of possible solutions is reduced to consider those that are above certain values for the DC and the DS. 
To illustrate, we consider the situation in which such lower bounds are set to $0.45$ and $0.55$ for the DS and the DC, respectively.  In this case, the prototype tool is able to identify a set of $6$ Pareto optimal solutions by taking around $1.4$ seconds and then consumes around $5$ milliseconds to compute the associated values $x_{T_1}, ..., x_{T_5}$.  Finally, the tool computes the set of security controls associated to the $6$ Pareto optimal solutions in around a second by solving (a variant of) the SSP as explained above for the single Pareto optimal solution.  

\subsection{\textbf{Experimental Results}}
\label{subsec:experimental-results}
\begin{table}[!b]
\centering
\caption{Experimental results of Test 1. Legend: Reduction Factor, Computation Time is in Seconds (S), and the maximum Heap Size is in Gigabyte (GB).}
% PLEASE MAJID CHECK THE NUMBERS... I HAVE DONE THE MATH AND THE REDUCTION FACTOR AT LEAST FOR THE FIRST 4 LINES DOES NOT SEEM CORRECT... THE REDUCTION SEEMS MUCH MORE DRAMATIC \multirow{4}{*}{4}!!!!
\label{tab:experimental}
\footnotesize
\begin{tabular}{ccccccc}
\toprule
  \multirow{2}{*}{\begin{tabular}[c]{@{}c@{}}$|\mathcal{T}|$\end{tabular}} &
  \multirow{2}{*}{\begin{tabular}[c]{@{}c@{}} $|\mathcal{C}_T|$\end{tabular}} &
  \multicolumn{2}{c}{Solution Set Size} &
  \multirow{2}{*}{\begin{tabular}[c]{@{}c@{}}Reduction \\ Factor \end{tabular}} &
  \multirow{2}{*}{\begin{tabular}[c]{@{}c@{}}Computation \\ Time (S)\end{tabular}} &
  \multirow{2}{*}{\begin{tabular}[c]{@{}c@{}} Heap Size \\(GB)\end{tabular}} \\ 
  \cmidrule(lr){3-4}      &    & $\Pi_{T\in \mathcal{T}}( k^{|\mathcal{C}_T|}-1)$ & $\Pi_{T\in \mathcal{T}} |{X}_T|$   &            &         &            \\ \midrule
5 &  20 & $32,768\cdot 10^5$ %$(3^4-1)^5$
& $32,768$      & $10^5$  & 0.312     & $\sim$0.25 \\
                  6 & 24 & $262,144\cdot 10^6$ %$(3^4-1)^6$
& $262,144$     & $10^6$    & 1.2    & $\sim$1.5  \\
                  7 & 28 & $2,097,152\cdot 10^7$ %$(3^4-1)^7$          
& $2,097,152$   & $10^7$   & 9.7    & $\sim$12   \\ 
                  8 &32 & $16,777,216\cdot 10^8$ %$(3^4-1)^8$          
& $16,777,216$  & $10^8$ & 237  & $\sim$29   \\ \midrule
                  5 &25 & $\sim 8.29\cdot 10^{11}$ %$(3^5-1)^5$           
& 100,000     & $\sim 8.29\cdot 10^6$    & 0.626     & $\sim$0.5  \\
                  6 &30  & $\sim 2.01\cdot 10^{14}$ %$(3^5-1)^6$          
& 1,000,000   & $\sim 2.01\cdot 10^8$  & 3.7    & $\sim$9    \\
                  7 &35  & $\sim 4.86\cdot 10^{16}$ %$(3^5-1)^7$           
& 10,000,000  & $\sim 4.86 \cdot 10^9$   & 105  & $\sim$28   \\
                  8 & 40 & $\sim 1.17\cdot 10^{19}$ %$(3^5-1)^8$           
& 100,000,000 & $\sim 1.17\cdot 10^{11}$& 2,787 & $\sim$416  \\ \bottomrule
\end{tabular}
\end{table}

This section undertakes some experimental evaluations to examine the scalability of proposed methodology through the implemented tool. Hence, we present two test cases to assess the computational time and resources in the following.  Since the instances of the variant of the SSP required to enumerate all possible mitigation mappings corresponding to each Pareto optimal solution of the form $\langle x_T \rangle_{T\in \mathcal{T}}$ are typically small, their solution does not consume a relevant amount of resources (both time and memory) and thus we disregard this activity in the discussion below.

\subsubsection{\textbf{Test 1: upfront computation of feasible solutions}}
The goal of the first set of tests is to evaluate the strategy of computing the set of feasible solutions upfront and then identify those that are Pareto optimal.  The idea is to understand the time and memory occupation required to do this while increasing the number of threats and the number of security controls per threat.  We consider two stakeholders (i.e.\ $|\mathcal{S}|=2$), the protection criteria $\mathcal{P}$ are the same as those in Table~\ref{tab:threat-preference-association}, the number of protection goals are $6$ as those introduced in Section~\ref{subsec:il-refinement}, an increasing number $|\mathcal{T}|=5, 6, 7, 8$ of threats, and a number $q=4,5$ of security control associated with each threat so that $|\mathcal{C}_T|=q*5, q*6, q*7, q*8$.  For each one of these configurations, we measure the time (in seconds) and the memory occupation (in GB of heap) taken to compute the entire set of feasible solutions when running our prototype on a cluster with a CPU of $3.2$ GHz and 500 GB of RAM.  We do not include the time to identify the Pareto Optimal solutions as the resource consumption for computing the feasible set of solutions (see the last two columns of Table~\ref{tab:experimental}) clearly shows the exponential behavior for both computation time and memory occupation despite the dramatic reduction in the search space (consider the values in the column Reduction Factor) obtained by using the approach of solving with respect to risk residues in place of mitigation mappings discussed at the end of Section~\ref{subsec:problem}.  
\subsubsection{\textbf{Test 2: interleaving the computation of feasible and optimal solutions}}
The first test set clearly shows that the upfront computation of the whole set of feasible solutions %followed by the selection of the Pareto Optimal solutions 
does not scale.  For this reason, we designed a different approach whereby the two activities are interleaved by computing non-overlapping sub-sets of the feasible solutions and then identify those that are Pareto Optimal.  As already observed, this can be done in different ways and we propose two strategies both parameterized by the size $d$ of the sub-set of feasible solutions that are being considered.
\begin{itemize}
    \item In the first strategy, we collect the Pareto Optimal solutions identified in each sub-set with cardinality $d$ of the set of feasible solutions in a list $\ell$ and once the entire set of feasible solutions has been covered, the list $\ell$ is processed to extract the final set of Pareto Optimal solutions. 
    \item The second strategy is similar to the previous one except for the fact that the content of the list $\ell$ of Pareto Optimal solutions for a given sub-set of the set of feasible solutions is added to the next sub-set of feasible solutions to be considered so that, when considering the last sub-set, we identify the final set of Pareto Optimal solutions.  
\end{itemize} 
To study the scalability in terms of resource consumption of these two strategies, we define a second test set with the same parameters of the previous one except for $|\mathcal{T}|=6,7,8,9$ and the number of security controls $q$ associated to each threat is $4$.  We consider increasing values of $d=8^h$ for $h=1, 2, 3, 4, 5, 6$ to understand how the cardinality of the sub-set of the feasible solutions affect performances.  
As for the previous test set, we measure the timing (in seconds) and the heap occupation (in MB) with a time out (T/O) of $3$ hours.  As the results---obtained on a personal computer with a CPU of $1.90$ GHz and $16$ GB of RAM---in Table~\ref{tab:experimental2} shows, the scalability is much improved with respect to the results of the first test above, regardless of the strategy adopted to identify the Pareto Optimal solutions. 
It is worth noticing that for this test set we consider a less powerful computer and include the computation for identifying the Pareto Optimal solutions. 
Although there is no clear winner between the two strategies described above, a closer analysis of the results in Table~\ref{tab:experimental2} shows that the second strategy is better than the first one in most cases and in particular for larger instances of the MSRMP; for example, consider the test case with $8$ threats and $d=8$, the computation time and the maximum heap space used by the first strategy are $2,098.4$ seconds and $1,282$ MB, whereas those used by the second strategy are $68.7$ seconds and $256$ MB.  We observe that setting an appropriate value for the parameter $d$ (neither too small nor too large) seems to be crucial for the timing behavior of first strategy while the second strategy seems to be much less independent; unsurprisingly, for the memory occupation, larger values of $d$ corresponds to larger heap sizes but much less than those of the first test set (notice that the numbers in Table~\ref{tab:experimental} are in GB whereas those in Table~\ref{tab:experimental2} are in MB).

\begin{sidewaystable}
\footnotesize
\centering
\caption{Experimental results based on the two defined strategies.}
\label{tab:experimental2}
\begin{tabular}{ccccccccc}
\toprule
\multirow{2}{*}{\begin{tabular}[c]{@{}c@{}}Test Case\\\end{tabular}} &
  \multirow{2}{*}{\begin{tabular}[c]{@{}c@{}} $|\mathcal{T}|$ \\  \end{tabular}}&
  \multirow{2}{*}{\begin{tabular}[c]{@{}c@{}} $|\mathcal{C_T}|$\end{tabular}} &
  \multicolumn{6}{c}{\multirow{3}{*}{\begin{tabular}[c]{@{}c@{}}Computation Time (Second) and RAM Heap Size (Megabyte)\\ \\\\\end{tabular}}} \\ \cmidrule(lr){4-9}
 &
   &
   &
  d=8 &
  d=64 &
  d=512 &
  d=4,096 &
  d=32,768 &
  d=262,144 \\ \hline
\multirow{4}{*}{Strategy 1} &
  6 &
  24 &
  4.7(S)   ,  308(MB) &
  2.5(S), 256(MB) &
  2.7(S), 256(MB) &
  3.7(S),    320(MB) &
  11.3(S),  499(MB) &
  6.8(S),  2,422(MB) \\
 &
  7 & 28
   &
  95.5(S),  986(MB) &
  8.1(S),  382(MB) &
  8.7(S), 256(MB) &
  10.7(S),  459(MB) &
  15.8(S),  900(MB) &
  357(S),  2,509(MB) \\
 &
  8 &32
   &
  2,098.4(S),  1,282(MB) &
  71(S),  308(MB) &
  52(S),  308(MB) &
  62.5(S),  497(MB) &
  159.5(S),  1,004(MB) &
  317.5(S),  3,500(MB) \\
 &
  9 &36
   &
  $T/O$ &
  7,346.7(S), 533(MB) &
  541.3(S),  308(MB) &
  560(S),  522(MB) &
  575.9(S),  1575(MB) &
  5,124.9(S),  3,812(MB) \\ \hline
\multirow{4}{*}{Strategy 2} &
  6 &
  24 &
  2.5(S), 256(MB) &
  4.5(S), 256(MB) &
  2.7(S), 256(MB) &
  3.9(S),  308(MB) &
  5.6(S),  826(MB) &
  7.2(S),  2,405(MB) \\
 &
  7 &28
   &
  10.7(S), 256(MB) &
  10.7(S), 256(MB) &
  12.7(S), 256(MB) &
  8.9(S),  525(MB) &
  10.9(S),  1,037(MB) &
  60.9(S),  2,471(MB) \\
 &
  8 &32
   &
  68.7(S),  256(MB) &
  83.3(S), 256(MB) &
  70.9(S), 256(MB) &
  58.3(S), 256(MB) &
  64.7(S), 1,186(MB) &
  108.9(S), 4,066(MB) \\
 &
  9 &36
   &
  567.3(S), 256(MB) &
  557.2(S), 308(MB) &
  507.6(S), 308(MB) &
  553.7(S), 408(MB) &
  555.8(S), 1,513(MB) &
  934(S), 4,066(MB)

 \\ \bottomrule
\end{tabular}
\end{sidewaystable}

\paragraph{Discussion on experiments}  There are two main lessons learned from the experiments discussed above.  First, the transformation of the original MSRMP (\ref{eq:msrmp}) over $\langle \mu_T\rangle_{T\in \mathcal{T}}$ into the one (\ref{eq:msrmp2}) over the $\langle x_T\rangle_{T\in \mathcal{T}}$ allows for a substantial reduction of the search space.  To see this, consider the Reduction Factor in Table \ref{tab:experimental}.  Second, considering the family $\mathcal{C}_{T\in \mathcal{T}}$ of controls associated to each threat $T\in \mathcal{T}$ is crucial, in practice, to reduce the search space of the problem of transforming back a solution $\langle x_T^*\rangle_{T\in \mathcal{T}}$ of (\ref{eq:msrmp2}) into the set $\{\langle \mu_T\rangle_{T\in \mathcal{T}}\}$ of associated mitigation mappings of the original MSRMP (\ref{eq:msrmp}).  This is so because the cardinality of $\mathcal{C}_T$ is usually low for each $T\in \mathcal{T}$ so that, despite the exponential complexity as discussed at the end of Section \ref{subsec:problem}, the time and memory consumption are reasonable in practice.  
\section{Related Work}\label{sec:relatedwork}

The most closely related work is~\cite{mollaeefar2020} that considers a similar---albeit simpler---optimization problem allowing for finding the best possible solutions among a (finite and small) set of possible RMPs.  Indeed, such solutions are not guaranteed to be Pareto optimal as those of the MSRMP considered in this paper.  Additionally in~\cite{mollaeefar2020}, no methodology to identify the set of possible RMPs is provided whereas this work provides a structured methodology for the definition of the whole set of RMPs via the notion of MSRMP.

In the scope of information security, a wide range of risk assessment approaches have been proposed by standard institutes and organizations like NIST SP 800-30 (NIST, 2012), ISO/IEC 27005 (2011), etc. Regardless of the particular processes each of these security risk assessment approaches have, they all point out to the risk as an unexpected incident that would damage business assets, either tangible (e.g., organization’s infrastructures) or intangible (e.g., organization’s services). 
The ultimate goal of an information security program based on risk management is to augment the organization’s output (product and service) while simultaneously limiting the unexpected adverse outcomes generated by potential risks. These methodologies have several limitations when intending to use them to analyze the risk from multi-stakeholder perspectives. Apart from that, for example, these frameworks are restricted in terms of what risks are related to data subjects and how to evaluate them, which is requested by the law.
Numerous methodologies and frameworks in the context of privacy impact assessment (PIA) have proposed, such as legal frameworks for data protection authorities in several countries~\cite{ref_article8,ref_article9,ref_article10},
as well as academic researchers~\cite{ref_article4,ref_article12,ref_article13,ref_article14}, and for specific purposes like PIA for RFID and Smart Grids~\cite{ref_article15,ref_article16}. 
By using PIA methodologies, technical and organizational privacy threats can be identified to select proper privacy controls. The classical risk assessments look at the risk from organizations’ view, often considered the security targets that must be protected. However, GDPR obligates controllers to conduct a Data Protection Impact Assessment (DPIA), stipulated by article 35. DPIA is required when a system relies on personal data processing because of the variety of privacy breaches that could arise. PIA approaches help to capture these breaches in the early stage and avoid them or reduce their impact by using appropriate measures~\cite{ref_article17,ref_article18}. There are shared traits between a privacy risk analysis and a security risk analysis. Security risks are those risks that arise from the loss of confidentiality, integrity, or availability of information or information systems and reflect the potential adverse impacts to organizational operations (i.e., mission, functions, image, or reputation), organizational assets, individuals, other organizations, and the Nations~\cite{ref_article20}. However, privacy is a more sophisticated, multifaceted concept aiming at protecting people and ruled by the laws~\cite{ref_article20}. Indeed, the GDPR concerns about the risk to data subject’s rights and freedoms. There are a few approaches that have defined risk impact criteria for different stakeholders. For instance, in~\cite{ref_article14}, the authors provide a seven-step approach to PIA, which is adopted from the NIST security risk assessment process (NIST, 2002). They have declared that privacy risk shall be assessed from both data subjects and system perspective. Similarly, Iwaya et al.~\cite{ref_article5} propose a privacy risk assessment by considering both perspectives. Their approach is based on the PIA methodology proposed by~\cite{ref_article14} in the case of mobile health data collection system, which is proposed a systematic identification and evaluation of privacy risks. In the context of cloud computing, in~\cite{ref_article32} a security risk assessment framework proposed that can enable cloud service providers to assess security risks in the cloud computing environment and allow cloud clients with different risk perspectives to contribute to risk assessment. 
In analyzing the conflict of interest between the risk owner and the risk actors in~~\cite{ref_article34} authors proposed a conflicting incentives risk analysis (CIRA) method in which risks are modeled in terms of conflicting incentives. CIRA’s goal is to provide an approach in which the input parameters can be audited more easily. Nevertheless, these approaches do not provide a quantification risk assessment to see the diverge of risk exposures from different perspectives.

% There a lot of risk assessment approaches which consider multi-criteria to calculate risk exposure. In~\cite{ref_article26} risk analysis is modeled as a Multi-Criteria Decision Making (MCDM) problem in which experts express their preferences for each risk, over two traditional criteria: probability and impact. 
% A risk-based decision framework~\cite{ref_article25} is proposed for cybersecurity strategy prioritization. 

\section{Conclusions and Future work}\label{sec:conclusion}

We have introduced the Multi-Stakeholder Risk Minimization Problem (MS\-RMP) to assist in the definition of the best (with respect to all the stakeholders involved in the system) Risk Management Policies (RMPs)---as an appropriate set of security controls to mitigate the identified set of threats---in the fundamental step of selecting mitigations for risk management.  
We have formalized the MSRMP as a multi-objective optimization problem that can be solved by using state-of-the-art techniques for Pareto Optimality.  On top of such techniques, we have proposed a semi-automated approach to define and solve instances of the MSRMP.  We have also discussed strategies to reduce the large search space resulting from real instances of the MSRMP.  We have illustrated the main notions of our approach on a simple yet representative running example. An implementation of the proposed approach has allowed us to perform an experimental evaluation whose results confirm the practical viability of the proposed approach. 

As future work, we consider three possibilities.  First, we plan to further validate the flexibility of our approach by integrating it with a methodology for the risk evaluation of identity proofing solutions introduced in~\cite{pernpruner2021}.  In that work, the authors present a framework composed to analyze the risks of enrollment solutions at the design time.  In particular, they focus on associating security controls with threats deriving from a set of attackers, so to reduce risks at an acceptable level while guaranteeing usability and economy.  However, it is left open the problem of determining the optimal set of mitigations, and this is the reason for which the approach presented in this work becomes an interesting complement.  The second (medium term) possibility for future work is to identify a comprehensive baseline of controls (such as the one in the Risk Management Framework of NIST\footnote{\url{https://csrc.nist.gov/Projects/risk-management/about-rmf/select-step}}) and provide an approach to tailor it to the use case scenario under consideration in order to lower the barrier of adoption of the approach proposed here by addressing the intricacies of evaluating the trade-offs of security controls including costs and skills required.  The third (and longer term) line of future work is to investigate how it is possible to smoothly combine the  approach proposed in this work with available methodologies for risk management (e.g., STRIDE).

\bibliography{ref}
\end{document}